\documentclass[amsmath,amssymb,reprint,nofootinbib,round]{revtex4-2}
\usepackage{mathptmx}
\usepackage[T1]{fontenc}
\usepackage[utf8]{inputenc}
\usepackage{xcolor}
\usepackage{amsmath}
\usepackage{graphicx}
\PassOptionsToPackage{normalem}{ulem}
\usepackage{ulem}
\usepackage[pdfusetitle,
 bookmarks=true,bookmarksnumbered=false,bookmarksopen=false,
 breaklinks=false,pdfborder={0 0 1},backref=false,colorlinks=false]
 {hyperref}

\makeatletter

\providecolor{lyxadded}{rgb}{0,0,1}
\providecolor{lyxdeleted}{rgb}{1,0,0}
\DeclareRobustCommand{\mklyxadded}[1]{\bgroup\color{lyxadded}{}#1\egroup}
\DeclareRobustCommand{\mklyxdeleted}[1]{\bgroup\color{lyxdeleted}\mklyxsout{#1}\egroup}
\DeclareRobustCommand{\mklyxsout}[1]{\ifx\\#1\else\sout{#1}\fi}
\DeclareRobustCommand{\lyxadded}[4][]{\texorpdfstring{\mklyxadded{#4}}{#4}}

\makeatother

\begin{document}
\title{Coupling a Fabry--Pérot Cavity to a Single-Mode Optical Fiber Using
a Metalens}
\author{Wance Wang}
\affiliation{Department of Physics, University of Maryland, College Park, Maryland,
USA}
\author{Wenqi Zhu}
\affiliation{Physical Measurement Laboratory, National Institute of Standards and
Technology, Gaithersburg, MD, USA}
\author{Amit Agrawal}
\affiliation{Department of Engineering, University of Cambridge, Cambridge, UK}
\author{Joseph W. Britton}
\affiliation{Department of Physics, University of Maryland, College Park, Maryland,
USA}
\affiliation{DEVCOM Army Research Laboratory, Adelphi, MD, USA}
\begin{abstract}
Efficient coupling of light from an optical cavity to a single-mode
fiber is required in a range of quantum technologies. In this work
we consider the coupling of a high-finesse macroscopic Fabry-Pérot
(FP) cavity to a single-mode fiber using a metalens. We perform sensitivity
analysis with respect to longitudinal and transverse misalignment
errors. We then detail a fiber-coupled cavity at $1650\text{ nm}$
using a monolithic cryo-compatible assembly incorporating a metalens.
\end{abstract}
\maketitle
\lyxadded{JWB}{Wed May 28 17:47:55 2025}{}

\section{Introduction}

Photon collection enables qubit readout and fluorescence imaging in
a variety of quantum platforms including neutral atoms \citep{Robens17ol,Cooper18prx,Browaeys20np},
trapped ions \citep{He21rosi,Carter24rosi,Guo24n}, color centers
\citep{Bucher19np} and quantum dots \citep{senellart-white2017nna}.
These systems can also generate single photons, a critical resource
for quantum networking \citep{Duan01n,Reiserer15rmp,Hermans22n,Nadlinger22n}.
High-NA free-space optics permit impressive collection efficiency
from point emitters (up to $4\pi\times0.3\text{ sr}$ \citep{Robens17ol})
but rely on very complex optical assemblies. An alternative approach
places the emitter in an optical cavity that enhances photon emission
into a well-defined spatial mode via the Purcell effect \citep{Reiserer15rmp,Takahashi20prl,Brekenfeld20np},
but shifts complexity from refractive optics to arguably more complex
in-vacuum alignment-sensitive optical cavities. The optimization of
populating a cavity mode by an atom-generated photon is well studied
\citep{goto-aoki2019praa,Gao23pra,Hughes23oe}. The related problem
of efficiently coupling the cavity to a single-mode (SM) optical fiber
also requires careful analysis.

\lyxadded{JWB}{Fri May 30 21:17:18 2025}{}

The leading approaches to build in-vacuum Fabry-Pérot (FP) cavities
rely on mirrors formed on fiber tips \citep{Hunger10njp} or macroscopic
formed-glass mirrors \citep{rempe-lalezari1992ola,jin-rakich2022oa}.
For fiber-cavities, mode matching has been explored in depth \citep{Hunger10njp,Gallego16apb,Bick16rosi,Pfeifer22apb}.
And, up to 90\% coupling efficiency was demonstrated using graded
index (GRIN) fiber elements \citep{Gulati17sr}. However, this technique
is not readily transferable to macroscopic cavities as it relies on
fiber fusion-splicing. We extend the range of techniques for in-vacuum
fiber coupling of macroscopic mirrors by exploring the use of metamaterial-based
lenses.

Metalenses \citep{Chen16rpp} are two-dimensional engineered nano-structures
that manipulate the phase of light similar to a refractive optic but
are vanishingly thin and quite versatile\lyxadded{y-wan}{Wed Jun  4 04:37:39 2025}{}.
For example, $0.9\text{ NA}$ metalenses have been demonstrated with
diffraction efficiency of $0.89$ and total efficiency (including
reflection and absorption) of $0.67$\citep{phan-fan2019lsaa}. In
AMO physics, metalenses have been successfully used to trap neutral
atoms \citep{Hsu22pq,Huang24oe,Chen24lpr}, collect fluorescence from
an atom \citep{Lim24q2ce2pq}, generate orbital angular momentum light
field \citep{Ren19nc,Zhou21lpr}, and form an FP cavity \citep{Ossiander23nc}.
These designs have successfully reduced the footprint and enhanced
the mechanical robustness in their respective applications.

In this manuscript we explore coupling of a macroscopic FP cavity
to an SM optical fiber using a metalens. We work with a geometry where
one cavity mirror is free-space coupled and the other is fiber coupled.
In this context we analyze the sensitivity of fiber-cavity coupling
efficiency to a variety of misalignment errors. We then build a monolithic
assembly to couple the TEM00 mode of a FP cavity at $1650\text{ nm}$
into a telecom SM fiber using a metalens. Finally, we discus the tradeoffs
of using metamaterials for beam forming in the context of single-photon
experiments.

\lyxadded{JWB}{Tue May 27 22:11:37 2025}{}

\section{Geometric Alignment Tolerance\protect\label{sec:Design}}

\lyxadded{JWB}{Tue Jun  3 22:09:32 2025}{}The conventional approach
is to put the cavity in-vacuum along with non-adjustable refractive
collimating optics \citep{Carter24rosi}. The collimated beam then
passes thru a vacuum window and is fiber-coupled using an asphere.
\lyxadded{JWB}{Tue Jun  3 21:40:18 2025}{} Such spatially extended
beam lines are sensitive to vibration and drift---especially in cryogenic
setups \citep{Fontana21rosi,Wipfli23rosi,Kumar23n}. Here, we aim
to extend the range of cavity-fiber coupling options by exploring
a compact in-vacuum assembly with the fiber tip and fiber mode matching
optics positioned close to the cavity. We present our design below
and carry out a tolerance analysis to guide fabrication and use of
the assembly.

\begin{figure}
\includegraphics[width=9cm]{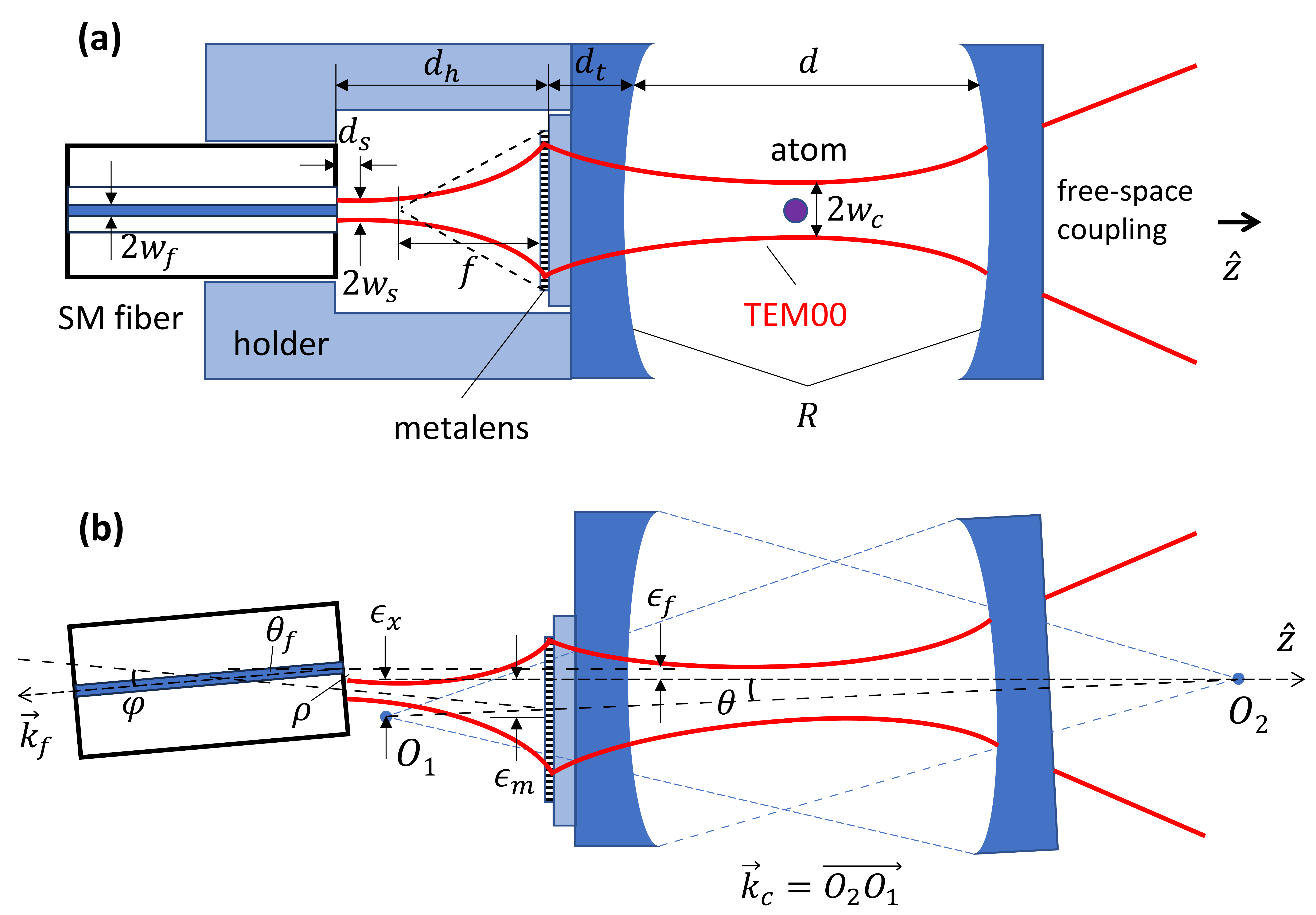}

\caption{\protect\label{fig:cross-section}Cross section of the optical assembly
consisting of a pair of mirrors forming a FP cavity, a metalens and
a SM fiber. (a) A diagram showing how longitudinal alignment errors
can produce a beam waist $w_{s}$ different from the fiber $w_{f}$
at a position $d_{s}$ away from the fiber tip. A mechanical holder
for the fiber is also shown. (b) A diagram showing transverse alignment
errors.}
\end{figure}

The goal of our optical assembly is to mode match a cavity TEM00 mode
to a single-mode fiber mode using a metalens. Our design of the assembly
is illustrated in Fig.$\:$\ref{fig:cross-section}(a). The cavity
has a length $d$, mirror radius of curvature $R$ and mirror thickness
$d_{t}$. The mirrors are mirror symmetric about the cavity waist
(radius $w_{c}$). The cavity mode is propagagted to the fiber tip
by the metalens (focal length $f$) at a position $d_{s}$ with a
waist $w_{s}$. An SM fiber (core mode field diameter $2w_{f}$) is
mechanically integrated with the left cavity mirror using a holder.\lyxadded{JWB}{Tue May 27 23:32:43 2025}{}
The distance from the metalens to the fiber tip is $d_{h}$.

We now vary $f$ and $d_{h}$ to maximize the coupling. The cavity
TEM00 mode waist is
\begin{equation}
w_{c}=\sqrt{\frac{\lambda}{\pi}}\left[\frac{d(2R-d)}{4}\right]^{1/4}
\end{equation}
\lyxadded{JWB}{Fri May 16 23:51:08 2025}{}The metalens lies distance
$d_{o}=d/2+d_{t}$ from the cavity waist. The lens equation is
\begin{equation}
\frac{1}{d_{o}+z_{R}^{2}/(d_{o}-f)}+\frac{1}{d_{h}-d_{s}}=\frac{1}{f}\label{eq:lens-eq}
\end{equation}
$z_{R}=\pi w_{c}^{2}/\lambda$ is the Rayleigh range. The transformed
beam waist is
\begin{align}
w_{s} & =\frac{1}{\sqrt{\left(1-d_{o}/f\right)^{2}+\left(z_{R}/f\right)^{2}}}w_{c}\label{eq:lens-mag}
\end{align}
. Ideal coupling is obtained by solving for $f$ and $d_{h}$ so that
$d_{s}=0$ and $w_{s}=w_{f}$.\lyxadded{JWB}{Wed May 28 18:58:33 2025}{}\lyxadded{y-wan}{Wed May 28 03:19:21 2025}{ }

\paragraph{Two cavity geometries\protect\label{par:geometry}}

To achieve high cooperativity with atomic quantum emitters, a small
cavity waist is desired \citep{Nguyen18pra,Krutyanskiy23prl,Periwal21n,Deist22prl,Shadmany25sa}.
Here we consider two example cavity geometries: one stable and one
near-concentric.
\begin{enumerate}
\item $R=50\text{ mm}$, $d=10\text{ mm}$: $w_{c}=86\text{ \ensuremath{\mu}m}$,
$f=1.056\text{ mm}$, $d_{h}=1.088\text{ mm}$. Stability parameter
$g_{1}g_{2}=(1-d/R)^{2}=0.64$.
\item $R=5\text{ mm}$, $d=9.98\text{ mm}$ (near-concentric): $w_{c}=10\text{ \ensuremath{\mu}m}$,
$f=3.310\text{ mm}$, $d_{h}=4.949\text{ mm}$. Stability parameter
$g_{1}g_{2}=(1-d/R)^{2}=0.99$.
\end{enumerate}
In both cases we assume $d_{t}=5\text{ mm}$,$\lambda=1550\text{ nm}$
and $2w_{f}=10.4\text{ \ensuremath{\mu}m}$ (typical for telecom
fiber \citep{Corning24Web}).

\subsection{Misalignment}

We identify errors that give rise to longitudinal and transverse deviations
that reduce fiber-cavity coupling. We then perform sensitivity analysis
to identify the most critical parameters.

\subsubsection{Longitudinal misalignment}

In this section we evaluate the sensitivity factor of various longitudinal
errors then calculate their impact on cavity-fiber coupling. Longitudinal
error includes variation in cavity length $d$, mirror thickness $d_{t}$,
length $d_{h}$, and focal length $f$. The result of all these errors
is a longitudinal shift of the transformed $d_{s}$ and waist size
$w_{s}$.

Doing the derivatives, we get the following sensitivities of $d_{s}$
and $w_{s}$.\begin{widetext}
\begin{align}
\frac{\partial d_{s}}{\partial d} & =\frac{2f^{2}(d_{t}-f)(R+d_{t}-f)}{\left[dR+2(d_{t}-f)(d+d_{t}-f)\right]^{2}}\nonumber \\
\frac{\partial d_{s}}{\partial d_{t}} & =\frac{2f^{2}\left[-dR+(d_{t}-f)^{2}+(d+d_{t}-f)^{2}\right]}{\left[dR+2(d_{t}-f)(d+d_{t}-f)\right]{}^{2}}\\
\frac{\partial d_{s}}{\partial d_{h}} & =1\nonumber \\
\frac{\partial d_{s}}{\partial f} & =-\frac{d^{2}\left[\left(R+2d_{t}-f\right)^{2}+f^{2}\right]+2dR\left[2d_{t}(d_{t}-f)-f^{2}\right]+4d_{t}(d_{t}-f)\left[dd_{t}+(d+d_{t})(d_{t}-f)\right]}{\left[dR+2(d_{t}-f)(d+d_{t}-f)\right]^{2}}\nonumber 
\end{align}
\begin{align}
\frac{\partial w_{s}}{\partial d} & =-\frac{dR^{2}+2dR(d_{t}-f)-2(R-d)(d_{t}-f)^{2}}{4d(2R-d)f^{2}}\left(\frac{w_{f}}{w_{c}}\right)^{2}w_{f}\nonumber \\
\frac{\partial w_{s}}{\partial d_{t}} & =-\frac{d+2d_{t}-2f}{2f^{2}}\left(\frac{w_{f}}{w_{c}}\right)^{2}w_{f}\\
\frac{\partial w_{s}}{\partial d_{h}} & =0\nonumber \\
\frac{\partial w_{s}}{\partial f} & =\frac{1}{f}w_{f}+\frac{d+2d_{t}-2f}{2f^{2}}\left(\frac{w_{f}}{w_{c}}\right)^{2}w_{f}\nonumber 
\end{align}
\end{widetext}Given error $\Delta d$, for example, the refracted
beam deviation is $\Delta d_{s}=\frac{\partial d_{s}}{\partial d}\Delta d$
and $\Delta w_{s}=\frac{\partial w_{s}}{\partial d}\Delta d$. 

The cavity-fiber coupling can be calculated from an overlap integral
between their mode shapes. We assume both are normalized Gaussians.
Recall that a Gaussian with waist $w_{0}$ located at $z=0$ has the
form
\begin{equation}
\boldsymbol{E}(x,y,z)=\frac{2}{\pi}\frac{1}{w(z)}\exp\left(-\frac{x^{2}+y^{2}}{w(z)^{2}}-ik\left(z+\frac{x^{2}+y^{2}}{2R(z)}-\psi(z)\right)\right)
\end{equation}
with beam radius $w(z)=w_{0}\sqrt{1+\left(z/z_{R}\right)^{2}}$, radius
of curvature $R(z)=z\left[1+\left(z_{R}/z\right)^{2}\right]$ and
Gouy phase $\psi(z)$. Here, the Rayleigh range is $z_{R}=\frac{1}{2}kw_{0}^{2}$
and $k=2\pi/\lambda$. The overlap integral is
\begin{equation}
\eta=\left|\int_{-\infty}^{\infty}\int_{-\infty}^{\infty}\boldsymbol{E}_{s}(x,y,0)\boldsymbol{E}_{f}^{*}(x,y,0)dxdy\right|^{2}
\end{equation}
$\boldsymbol{E}_{s}$ is the cavity TEM00 Gaussian mode imaged by
the metalens and $\boldsymbol{E}_{f}$ is the Gaussian mode with a
waist at the fiber tip (assume the fiber tip at $z=0$). Waist $w_{0}=w_{s}$
for $\boldsymbol{E}_{s}$ and $w_{0}=w_{f}$ for $\boldsymbol{E}_{f}$.

\begin{figure*}
\includegraphics[width=11.5cm]{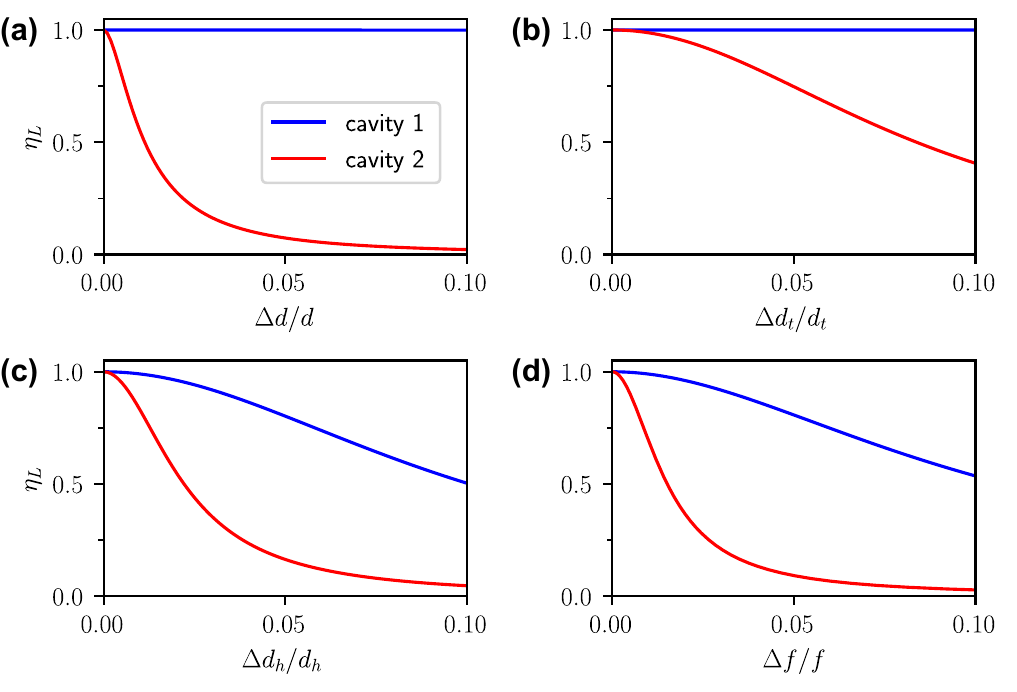}\caption{\protect\label{fig:longitudinal}(a-d) Coupling efficiency $\eta_{L}$
a function of $\Delta d$, $\Delta d_{t}$, $\Delta d_{h}$, and $\Delta f$
of the two cavity geometries. Values of $d,d_{t},d_{h},f$ for the
two geometries are listed in \ref{par:geometry}.}
\end{figure*}

Ref.$\:$\citep{Joyce84ao} relates errors $d_{s}+\Delta d_{s}$ and
$w_{s}+\Delta w_{s}$ to efficiency. 
\begin{equation}
\eta_{L}(\Delta d_{s},\Delta w_{s})=\frac{4}{\left(\frac{w_{s}+\Delta w_{s}}{w_{f}}+\frac{w_{f}}{w_{s}+\Delta w_{s}}\right)^{2}+\left[\frac{2}{k\left(w_{s}+\Delta w_{s}\right)w_{f}}\right]^{2}\left(\Delta d_{s}\right)^{2}}
\end{equation}
Given $w_{s}=w_{f}$ and $d_{s}=0$, $\eta_{L}(0,0)=1$. Any longitudinal
misalignment $p\in\{d,d_{t},d_{h},f\}$ results in coupling efficiency
$\eta_{L}\left(\frac{\partial d_{s}}{\partial p}\Delta p,\frac{\partial w_{s}}{\partial p}\Delta p\right)$\footnote{For cavity geometry 2, we change cavity length $d$ to $d-\Delta d$
to avoid pushing the cavity to instability.} (Fig.$\:$\ref{fig:longitudinal}).

For the near-concentric cavity efficiency drops to $50\%$ when any
one of the following errors occurs: $\Delta d=100\text{ \ensuremath{\mu}m}$,
$\Delta d_{t}=420\text{ \ensuremath{\mu}m}$, $\Delta d_{h}=110\text{ \ensuremath{\mu}m}$,
or $\Delta f=50\text{ \ensuremath{\mu}m}$. Let's check some routine
uses cases. When aligning the cavity it's routine to scan the cavity
length by 2 free spectral ranges (FSR) to check cavity spectrum, which
corresponds to $\Delta d=2\lambda=3.1\text{ \ensuremath{\mu}m}\ll100\text{ \ensuremath{\mu}m}$.
For the fused silica mirror and holder, it will not expand or contract
by $\Delta d_{t}/d_{t}$ and $\Delta d_{h}/d_{h}$ given any reasonable
temperature change. For a metalens manufactured by electron-beam lithography,
the focal length error is approximately $1\text{ \ensuremath{\mu}m}$\lyxadded{JWB}{Wed May 28 00:29:26 2025}{}.
Cavity 1 can tolerate much larger longitudinal error. Diffraction
and clipping loss are no problem as the metalens diameter is much
larger than the beam diameter at $d/2+d_{t}$.

\subsubsection{Transverse misalignment}

Transverse errors cause optical path deviations relative to the $\hat{z}$
optical axis. We define $\hat{z}$ as the axis passing through the
center of left mirror's circular substrate and the center of curvature
$O_{2}$ of the left mirror (see Fig.$\;$\ref{fig:cross-section}(b)).
First, if the center of curvature $O_{1}$ of the right mirror is
transversely displaced from $\hat{z}$ by $\epsilon_{x}$, the resulting
cavity axis $\vec{k}_{c}=\overrightarrow{O_{2}O_{1}}$ forms an angle
$\theta$ with $\hat{z}$. Second, the metalens axis could be offset
by $\epsilon_{m}$ relative to the center of the left mirror. Finally,
the fiber axis could be tilted by angle $\theta_{f}$ relative to
$\hat{z}$ or displaced by an offset $\epsilon_{f}$ left mirror .
All sources of transverse misalignment lead to a tilted image at the
fiber offset by $\rho$ relative to the fiber core and angle $\varphi$
relative to the fiber axis $\vec{k}_{f}$. Note that in our geometry
$\rho$ and $\varphi$ are dependent variables, that is, they are
not independently adjusted.

Using ray transfer matrix analysis, we calculate the sensitivities
of $\rho$ and $\varphi$ to transverse errors $\epsilon_{x}$, $\epsilon_{m}$
$\theta_{f}$ and $\epsilon_{f}$. We consider the worst case scenario
where all errors are additive. In this case,\begin{widetext}
\begin{align}
\begin{pmatrix}\rho\\
\varphi
\end{pmatrix} & =\frac{1}{2R-d}\begin{pmatrix}(R+d_{t})(1-\frac{d_{s}}{f})+d_{s}\\
\frac{f-R-d_{t}}{f}
\end{pmatrix}\epsilon_{x}+\begin{pmatrix}1-\frac{d_{s}}{f}\\
-\frac{1}{f}
\end{pmatrix}\epsilon_{m}+\begin{pmatrix}0\\
1
\end{pmatrix}\theta_{f}+\begin{pmatrix}1\\
0
\end{pmatrix}\epsilon_{f}
\end{align}
\end{widetext}The angle between $\vec{k_{c}}$ and $\hat{z}$ is
$\theta=\epsilon_{x}/(2R-d)$. We parameterize mirror-mirror misalignment
in terms of $\epsilon_{x}$ as it includes the $2R-d$ geometric factor.

\begin{figure*}
\includegraphics[width=11.5cm]{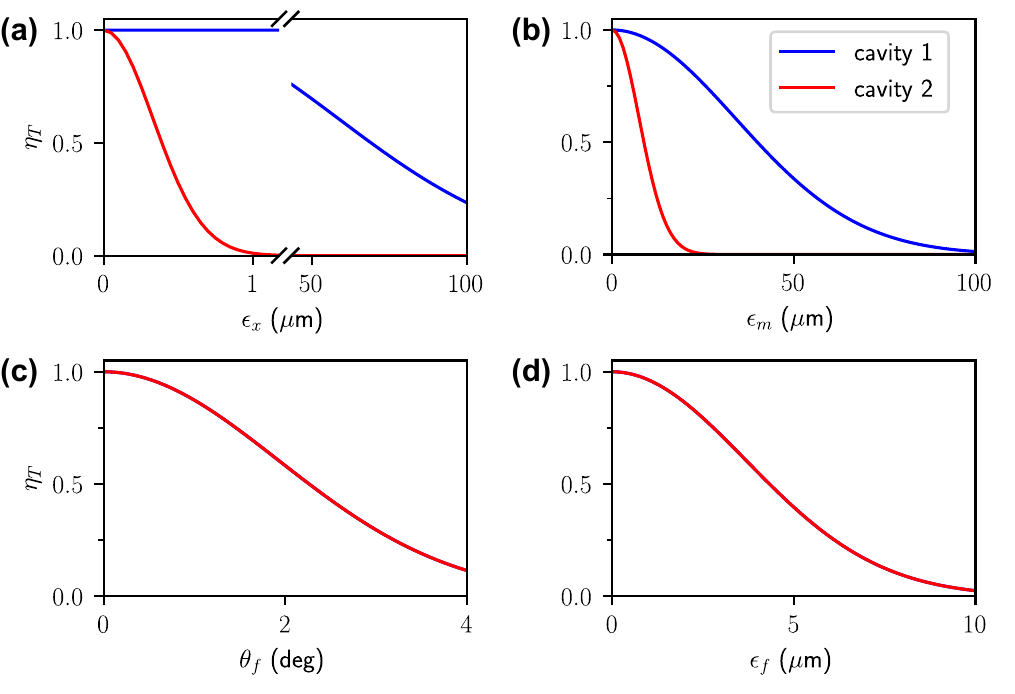}\caption{\protect\label{fig:transverse}(a-d) Coupling efficiency $\eta_{T}$
a function of $\epsilon_{x}$, $\epsilon_{m}$, $\theta_{f}$, and
$\epsilon_{f}$ of the two cavity geometries. In (c,d), traces of
two cavities overlap.}
\end{figure*}

Similarly, assuming no longitudinal error we calculate the efficiency
as a function of $\rho$ and $\varphi$ based on Ref.$\:$\citep{Joyce84ao},
\begin{equation}
\eta_{T}(\rho,\varphi)=\exp\left[-\left(\rho/w_{f}\right)^{2}-\left(\varphi kw_{f}\right)^{2}\right]
\end{equation}
. Any transverse misalignment $p\in\{\epsilon_{x},\epsilon_{m},\theta_{f},\epsilon_{f}\}$
results in a reduced efficiency $\eta_{T}\left(\frac{\partial\rho}{\partial p}\Delta p,\frac{\partial\varphi}{\partial p}\Delta p\right)$(see
Fig$\:$\ref{fig:transverse}).

For the near-concentric cavity (cavity 2) the efficiency drops to
$50\%$ when any one of the following errors occurs: $\epsilon_{x}=0.4\text{ \ensuremath{\mu}m}$
(corresponds to $\theta=1.1\text{ deg}$), $\epsilon_{m}=8.5\text{ \ensuremath{\mu}m}$,
$\theta_{f}=2.2\text{ deg}$, or $\epsilon_{f}=50\text{ \ensuremath{\mu}m}$.
The transverse errors $\epsilon_{x}$ and $\epsilon_{m}$ are particularly
consequential to coupling. The tight constraint on $\epsilon_{x}$
is an inherent challenge of near-concentric cavities. If the mechanical
assembly permits tuning of $\theta_{f}$ and offset $\epsilon_{f}$,
they can be used to compensate for errors in $\epsilon_{x}$ and $\epsilon_{m}$.
In contrast, the stable cavity (cavity 1) is much more robust to errors.\lyxadded{JWB}{Tue Jun  3 21:45:55 2025}{}\lyxadded{JWB}{Tue Jun  3 21:45:55 2025}{¶}

\section{Experimental Setup}

\begin{figure*}
\includegraphics[width=18cm]{Fig4_integration_1}

\includegraphics[width=8.5cm]{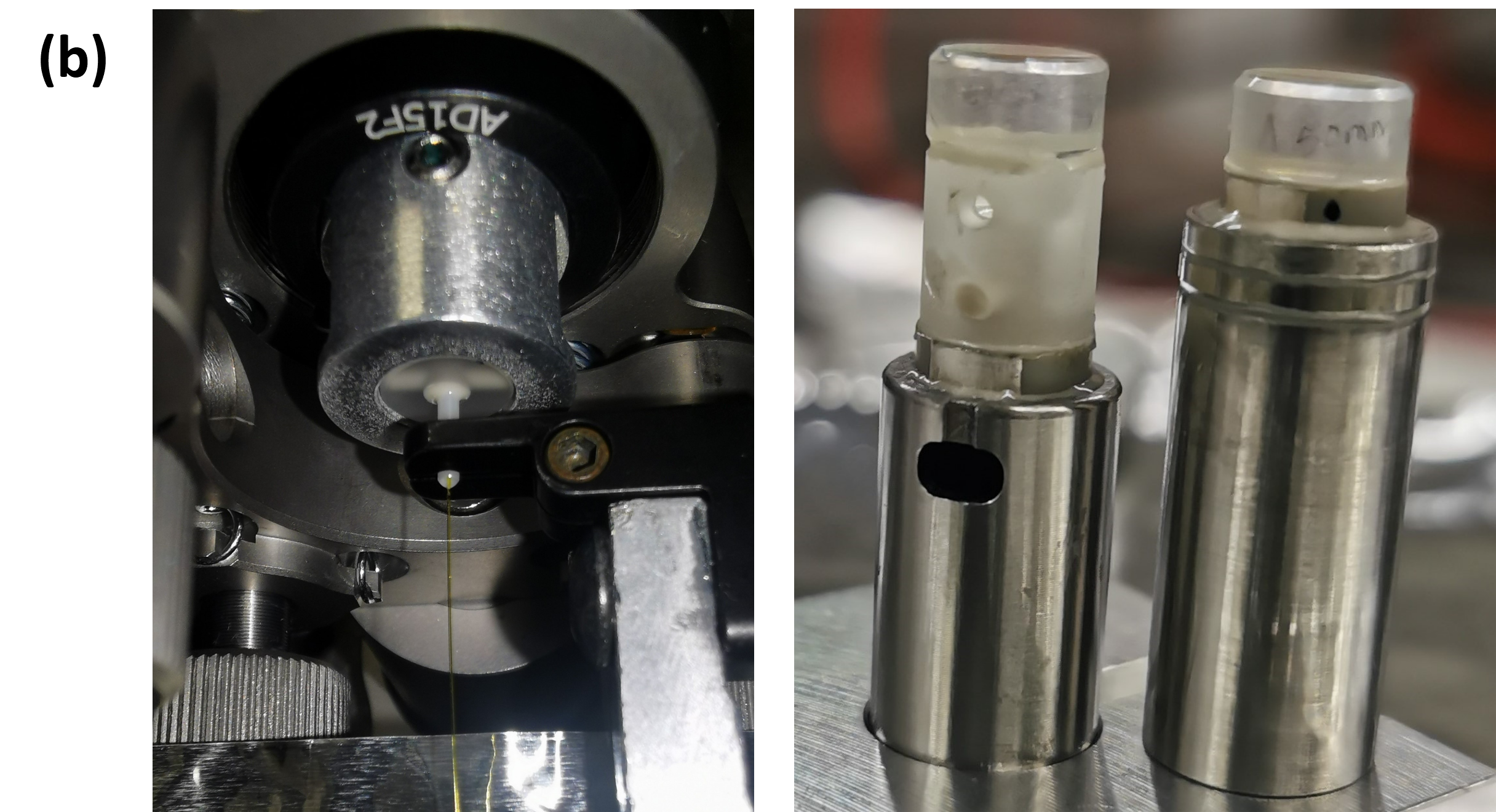}

\caption{\protect\label{fig:steps}(a) Illustration of the cavity-fiber integration
steps. (b) Left photo: gluing fiber in a ferrule to the holder. Right
photo: the finished assembly; left one is the cavity--metalens--fiber
assembly (the fiber extends below the metal jig), and the right one
is the free-space cavity mirror. They are both glued to metal tube
mounts.}
\end{figure*}

In this section we describe a monolithic optical assembly that combines
mirror, metalens, spacers, fiber and fiber mechanical support. It
also details alignment of the assembly in the context of an FP cavity
and our characterization of its coupling. We choose a particular cavity
geometry based on our atomic system and the constraints of a preexisting
ion trap. It closely resembles geometry 1.

Our cavity geometry has $R=50\text{ mm}$ and $d=13\text{ mm}$, with
optical coatings yielding a finesse of $\sim20,000$ at $1650\text{ nm}$,
similar to stable cavity geometry 1. We calculate that for optimal
mode matching $f=1.161\text{ mm}$ and $d_{h}=1.196\text{ mm}$.

The metalens for this experiment is constructed from $1.2\text{ \ensuremath{\mu}m}$-tall
polycrystalline Si nanopillars ($n=3.47$) patterned by electron
beam lithography and fluorine-based reactive ion etch. The nanopillar
array is fabricated on a $500\text{ \ensuremath{\mu}m}$-thick fused-silica
substrate ($n=1.44$) which is bonded to the AR-coated cavity mirror
using UV-cured Norland NOA-61 ($n=1.54$) \citep{no-endorsement}.
The metalens is designed for a convex focal length of $f=1.16\text{ mm}$
at $1650\text{ nm}$. At this wavelength the metalens transmission
is $0.92$ and the focusing efficiency is $0.61$\footnote{The fraction of light focused to a spot by the meta lens is called
the focusing efficiency; the balance of light passes straight thru
the metalens. With optimization, the focusing efficiency can be $0.90$\citep{phan-fan2019lsaa,EgedeJohansen24cp}
and may improve over time as metalens design and fabrication improves.}.

Figure$\:$\ref{fig:steps}(a) illustrates the assembly steps. To
mode match the TEM00 mode, the cavity must be well-defined and aligned.
In step (1), two optical mounts were aligned to a laser beam by maximizing
the transmitted power through two irises so the two mounts are simultaneously
concentric and perpendicular to the beam\lyxadded{JWB}{Wed May 28 15:07:20 2025}{}.
Cavity mirrors were installed in step (2) and we fine tuned one of
the mounts to align the cavity to the beam. In step (3), the left
mirror was uninstalled and placed horizontally under a stereoscope.
The metalens substrate was panned while visually checking its concentricity
relative to the mirror peripheral. We applied an optically transparent
epoxy A \citep{Norland24Web,no-endorsement} between the metalens
substrate and the mirror. The metalens was adjusted to its final position
and then fixed in place by UV curing the epoxy. In step (4), we tuned
fiber angle $\theta_{f}$ and offset $\epsilon_{f}$ to compensate
any offset $\epsilon_{m}$ introduced in step (3). The fiber, mounted
in a ferrule, was held in a stage with 5 degrees of freedom. Here,
the left mirror was re-installed and the right mirror was removed.
We adjusted the fiber to couple the transmitted light. Finally, in
step (5), we glued the holder to the outer rim of the left mirror.
The fiber was repositioned to maximize coupling, and the ferrule was
fixed in place by potting it inside the holder with epoxy B \citep{Masterbond24Web,no-endorsement}.
The assembly formed a monolithic part after epoxy curing.\lyxadded{JWB}{Wed May 28 15:10:13 2025}{}

During the assembly process we performed several diagnostic measurements.
In step (2), we scanned the laser frequency across the full cavity
FSR and observed a contrast ratio $>40:1$ between TEM00 and higher-order
transverse modes. This confirmed that the input beam closely approximated
the TEM00, allowing us in step (4) to use it for fiber coupling without
tuning the laser to cavity resonance. We also verified that uninstalling
and re-installing the mirrors did not degrade alignment. In step
(3), we constrain $\epsilon_{m}<200\text{ \ensuremath{\mu}m}$ (1/10
of the metalens diameter). Although this accuracy is poor, we compensate
in step (4) by tuning $\theta_{f}$ and $\epsilon_{f}$. We achieved
a relative fiber coupling efficiency of $94\%$ in step (4)\footnote{The relative fiber coupling efficiency excludes metalens transmission
losses and refraction efficiency.}, excluding cavity mirror transmission. Additionally, we observed
that intentionally displacing the right mirror by $\epsilon_{x}\approx100\text{ \ensuremath{\mu}m}$
reduced TEM00 fiber-coupled power by $50\%$, generally consistent
with the result in Fig.$\:$\ref{fig:transverse}(a).

During step (5) we found the the coupling efficiency dropped to $22\%$
due to epoxy shrinkage during epoxy curing. We tried to compensate
by adjusting the ferrule position but its response was sluggish after
the epoxy became viscous. We anticipate that greater vigilance and
active alignment during curing will avoid this problem. Figure$\:$\ref{fig:steps}(b)
shows the ferrule potted inside the holder and the appearance of the
completed assembly.

\section{Conclusion and Future}

We analyzed the impact of various misalignments on coupling a FP TEM00
mode to a single-mode fiber using a metalens. Our sensitivity and
coupling efficiency calculations reveal that transverse misalignment
is more critical than longitudinal ones. Our work complements recent
atom-cavity coupling analysis \citep{Gao23pra}. We then built and
tested a compact monolithic assembly. While the final coupling efficiency
was limited, the $94\%$ relative fiber coupling efficiency in step
(4) is representative of the excellent performance metameterial optics.

\section*{Bibliography}

\bibliographystyle{apsrev4-2}
\nocite{*}
\bibliography{2025metalens}

\begin{thebibliography}{63}%
\makeatletter
\providecommand \@ifxundefined [1]{%
 \@ifx{#1\undefined}
}%
\providecommand \@ifnum [1]{%
 \ifnum #1\expandafter \@firstoftwo
 \else \expandafter \@secondoftwo
 \fi
}%
\providecommand \@ifx [1]{%
 \ifx #1\expandafter \@firstoftwo
 \else \expandafter \@secondoftwo
 \fi
}%
\providecommand \natexlab [1]{#1}%
\providecommand \enquote  [1]{``#1''}%
\providecommand \bibnamefont  [1]{#1}%
\providecommand \bibfnamefont [1]{#1}%
\providecommand \citenamefont [1]{#1}%
\providecommand \href@noop [0]{\@secondoftwo}%
\providecommand \href [0]{\begingroup \@sanitize@url \@href}%
\providecommand \@href[1]{\@@startlink{#1}\@@href}%
\providecommand \@@href[1]{\endgroup#1\@@endlink}%
\providecommand \@sanitize@url [0]{\catcode `\\12\catcode `\$12\catcode
  `\&12\catcode `\#12\catcode `\^12\catcode `\_12\catcode `\%12\relax}%
\providecommand \@@startlink[1]{}%
\providecommand \@@endlink[0]{}%
\providecommand \url  [0]{\begingroup\@sanitize@url \@url }%
\providecommand \@url [1]{\endgroup\@href {#1}{\urlprefix }}%
\providecommand \urlprefix  [0]{URL }%
\providecommand \Eprint [0]{\href }%
\providecommand \doibase [0]{https://doi.org/}%
\providecommand \selectlanguage [0]{\@gobble}%
\providecommand \bibinfo  [0]{\@secondoftwo}%
\providecommand \bibfield  [0]{\@secondoftwo}%
\providecommand \translation [1]{[#1]}%
\providecommand \BibitemOpen [0]{}%
\providecommand \bibitemStop [0]{}%
\providecommand \bibitemNoStop [0]{.\EOS\space}%
\providecommand \EOS [0]{\spacefactor3000\relax}%
\providecommand \BibitemShut  [1]{\csname bibitem#1\endcsname}%
\let\auto@bib@innerbib\@empty
\bibitem [{\citenamefont {Robens}\ \emph {et~al.}(2017)\citenamefont {Robens},
  \citenamefont {Brakhane}, \citenamefont {Alt}, \citenamefont {Klei{\ss}ler},
  \citenamefont {Meschede}, \citenamefont {Moon}, \citenamefont {Ramola},\ and\
  \citenamefont {Alberti}}]{Robens17ol}%
  \BibitemOpen
  \bibfield  {author} {\bibinfo {author} {\bibfnamefont {C.}~\bibnamefont
  {Robens}}, \bibinfo {author} {\bibfnamefont {S.}~\bibnamefont {Brakhane}},
  \bibinfo {author} {\bibfnamefont {W.}~\bibnamefont {Alt}}, \bibinfo {author}
  {\bibfnamefont {F.}~\bibnamefont {Klei{\ss}ler}}, \bibinfo {author}
  {\bibfnamefont {D.}~\bibnamefont {Meschede}}, \bibinfo {author}
  {\bibfnamefont {G.}~\bibnamefont {Moon}}, \bibinfo {author} {\bibfnamefont
  {G.}~\bibnamefont {Ramola}},\ and\ \bibinfo {author} {\bibfnamefont
  {A.}~\bibnamefont {Alberti}},\ }\href {https://doi.org/10.1364/OL.42.001043}
  {\bibfield  {journal} {\bibinfo  {journal} {Opt. Lett.}\ }\textbf {\bibinfo
  {volume} {42}},\ \bibinfo {pages} {1043} (\bibinfo {year}
  {2017})}\BibitemShut {NoStop}%
\bibitem [{\citenamefont {Cooper}\ \emph {et~al.}(2018)\citenamefont {Cooper},
  \citenamefont {Covey}, \citenamefont {Madjarov}, \citenamefont {Porsev},
  \citenamefont {Safronova},\ and\ \citenamefont {Endres}}]{Cooper18prx}%
  \BibitemOpen
  \bibfield  {author} {\bibinfo {author} {\bibfnamefont {A.}~\bibnamefont
  {Cooper}}, \bibinfo {author} {\bibfnamefont {J.~P.}\ \bibnamefont {Covey}},
  \bibinfo {author} {\bibfnamefont {I.~S.}\ \bibnamefont {Madjarov}}, \bibinfo
  {author} {\bibfnamefont {S.~G.}\ \bibnamefont {Porsev}}, \bibinfo {author}
  {\bibfnamefont {M.~S.}\ \bibnamefont {Safronova}},\ and\ \bibinfo {author}
  {\bibfnamefont {M.}~\bibnamefont {Endres}},\ }\href
  {https://doi.org/10.1103/PhysRevX.8.041055} {\bibfield  {journal} {\bibinfo
  {journal} {Phys. Rev. X}\ }\textbf {\bibinfo {volume} {8}},\ \bibinfo {pages}
  {041055} (\bibinfo {year} {2018})}\BibitemShut {NoStop}%
\bibitem [{\citenamefont {Browaeys}\ and\ \citenamefont
  {Lahaye}(2020)}]{Browaeys20np}%
  \BibitemOpen
  \bibfield  {author} {\bibinfo {author} {\bibfnamefont {A.}~\bibnamefont
  {Browaeys}}\ and\ \bibinfo {author} {\bibfnamefont {T.}~\bibnamefont
  {Lahaye}},\ }\href {https://doi.org/10.1038/s41567-019-0733-z} {\bibfield
  {journal} {\bibinfo  {journal} {Nat. Phys.}\ }\textbf {\bibinfo {volume}
  {16}},\ \bibinfo {pages} {132} (\bibinfo {year} {2020})}\BibitemShut
  {NoStop}%
\bibitem [{\citenamefont {He}\ \emph {et~al.}(2021)\citenamefont {He},
  \citenamefont {Cui}, \citenamefont {Li}, \citenamefont {Qian}, \citenamefont
  {Chen}, \citenamefont {Ai}, \citenamefont {Huang}, \citenamefont {Li},\ and\
  \citenamefont {Guo}}]{He21rosi}%
  \BibitemOpen
  \bibfield  {author} {\bibinfo {author} {\bibfnamefont {R.}~\bibnamefont
  {He}}, \bibinfo {author} {\bibfnamefont {J.-M.}\ \bibnamefont {Cui}},
  \bibinfo {author} {\bibfnamefont {R.-R.}\ \bibnamefont {Li}}, \bibinfo
  {author} {\bibfnamefont {Z.-H.}\ \bibnamefont {Qian}}, \bibinfo {author}
  {\bibfnamefont {Y.}~\bibnamefont {Chen}}, \bibinfo {author} {\bibfnamefont
  {M.-Z.}\ \bibnamefont {Ai}}, \bibinfo {author} {\bibfnamefont {Y.-F.}\
  \bibnamefont {Huang}}, \bibinfo {author} {\bibfnamefont {C.-F.}\ \bibnamefont
  {Li}},\ and\ \bibinfo {author} {\bibfnamefont {G.-C.}\ \bibnamefont {Guo}},\
  }\href {https://doi.org/10.1063/5.0043985} {\bibfield  {journal} {\bibinfo
  {journal} {Review of Scientific Instruments}\ }\textbf {\bibinfo {volume}
  {92}},\ \bibinfo {pages} {073201} (\bibinfo {year} {2021})}\BibitemShut
  {NoStop}%
\bibitem [{\citenamefont {Carter}\ \emph {et~al.}(2024)\citenamefont {Carter},
  \citenamefont {O'Reilly}, \citenamefont {Toh}, \citenamefont {Saha},
  \citenamefont {Shalaev}, \citenamefont {Goetting},\ and\ \citenamefont
  {Monroe}}]{Carter24rosi}%
  \BibitemOpen
  \bibfield  {author} {\bibinfo {author} {\bibfnamefont {A.~L.}\ \bibnamefont
  {Carter}}, \bibinfo {author} {\bibfnamefont {J.}~\bibnamefont {O'Reilly}},
  \bibinfo {author} {\bibfnamefont {G.}~\bibnamefont {Toh}}, \bibinfo {author}
  {\bibfnamefont {S.}~\bibnamefont {Saha}}, \bibinfo {author} {\bibfnamefont
  {M.}~\bibnamefont {Shalaev}}, \bibinfo {author} {\bibfnamefont
  {I.}~\bibnamefont {Goetting}},\ and\ \bibinfo {author} {\bibfnamefont
  {C.}~\bibnamefont {Monroe}},\ }\href {https://doi.org/10.1063/5.0180732}
  {\bibfield  {journal} {\bibinfo  {journal} {Review of Scientific
  Instruments}\ }\textbf {\bibinfo {volume} {95}},\ \bibinfo {pages} {033201}
  (\bibinfo {year} {2024})}\BibitemShut {NoStop}%
\bibitem [{\citenamefont {Guo}\ \emph {et~al.}(2024)\citenamefont {Guo},
  \citenamefont {Wu}, \citenamefont {Ye}, \citenamefont {Zhang}, \citenamefont
  {Lian}, \citenamefont {Yao}, \citenamefont {Wang}, \citenamefont {Yan},
  \citenamefont {Yi}, \citenamefont {Xu}, \citenamefont {Li}, \citenamefont
  {Hou}, \citenamefont {Xu}, \citenamefont {Guo}, \citenamefont {Zhang},
  \citenamefont {Qi}, \citenamefont {Zhou}, \citenamefont {He},\ and\
  \citenamefont {Duan}}]{Guo24n}%
  \BibitemOpen
  \bibfield  {author} {\bibinfo {author} {\bibfnamefont {S.-A.}\ \bibnamefont
  {Guo}}, \bibinfo {author} {\bibfnamefont {Y.-K.}\ \bibnamefont {Wu}},
  \bibinfo {author} {\bibfnamefont {J.}~\bibnamefont {Ye}}, \bibinfo {author}
  {\bibfnamefont {L.}~\bibnamefont {Zhang}}, \bibinfo {author} {\bibfnamefont
  {W.-Q.}\ \bibnamefont {Lian}}, \bibinfo {author} {\bibfnamefont
  {R.}~\bibnamefont {Yao}}, \bibinfo {author} {\bibfnamefont {Y.}~\bibnamefont
  {Wang}}, \bibinfo {author} {\bibfnamefont {R.-Y.}\ \bibnamefont {Yan}},
  \bibinfo {author} {\bibfnamefont {Y.-J.}\ \bibnamefont {Yi}}, \bibinfo
  {author} {\bibfnamefont {Y.-L.}\ \bibnamefont {Xu}}, \bibinfo {author}
  {\bibfnamefont {B.-W.}\ \bibnamefont {Li}}, \bibinfo {author} {\bibfnamefont
  {Y.-H.}\ \bibnamefont {Hou}}, \bibinfo {author} {\bibfnamefont {Y.-Z.}\
  \bibnamefont {Xu}}, \bibinfo {author} {\bibfnamefont {W.-X.}\ \bibnamefont
  {Guo}}, \bibinfo {author} {\bibfnamefont {C.}~\bibnamefont {Zhang}}, \bibinfo
  {author} {\bibfnamefont {B.-X.}\ \bibnamefont {Qi}}, \bibinfo {author}
  {\bibfnamefont {Z.-C.}\ \bibnamefont {Zhou}}, \bibinfo {author}
  {\bibfnamefont {L.}~\bibnamefont {He}},\ and\ \bibinfo {author}
  {\bibfnamefont {L.-M.}\ \bibnamefont {Duan}},\ }\href
  {https://doi.org/10.1038/s41586-024-07459-0} {\bibfield  {journal} {\bibinfo
  {journal} {Nature}\ }\textbf {\bibinfo {volume} {630}},\ \bibinfo {pages}
  {613} (\bibinfo {year} {2024})}\BibitemShut {NoStop}%
\bibitem [{\citenamefont {Bucher}\ \emph {et~al.}(2019)\citenamefont {Bucher},
  \citenamefont {Aude~Craik}, \citenamefont {Backlund}, \citenamefont {Turner},
  \citenamefont {Ben~Dor}, \citenamefont {Glenn},\ and\ \citenamefont
  {Walsworth}}]{Bucher19np}%
  \BibitemOpen
  \bibfield  {author} {\bibinfo {author} {\bibfnamefont {D.~B.}\ \bibnamefont
  {Bucher}}, \bibinfo {author} {\bibfnamefont {D.~P.~L.}\ \bibnamefont
  {Aude~Craik}}, \bibinfo {author} {\bibfnamefont {M.~P.}\ \bibnamefont
  {Backlund}}, \bibinfo {author} {\bibfnamefont {M.~J.}\ \bibnamefont
  {Turner}}, \bibinfo {author} {\bibfnamefont {O.}~\bibnamefont {Ben~Dor}},
  \bibinfo {author} {\bibfnamefont {D.~R.}\ \bibnamefont {Glenn}},\ and\
  \bibinfo {author} {\bibfnamefont {R.~L.}\ \bibnamefont {Walsworth}},\ }\href
  {https://doi.org/10.1038/s41596-019-0201-3} {\bibfield  {journal} {\bibinfo
  {journal} {Nat Protoc}\ }\textbf {\bibinfo {volume} {14}},\ \bibinfo {pages}
  {2707} (\bibinfo {year} {2019})}\BibitemShut {NoStop}%
\bibitem [{\citenamefont {Senellart}\ \emph {et~al.}(2017)\citenamefont
  {Senellart}, \citenamefont {Solomon},\ and\ \citenamefont
  {White}}]{senellart-white2017nna}%
  \BibitemOpen
  \bibfield  {author} {\bibinfo {author} {\bibfnamefont {P.}~\bibnamefont
  {Senellart}}, \bibinfo {author} {\bibfnamefont {G.}~\bibnamefont {Solomon}},\
  and\ \bibinfo {author} {\bibfnamefont {A.}~\bibnamefont {White}},\ }\href
  {https://doi.org/10.1038/nnano.2017.218} {\bibfield  {journal} {\bibinfo
  {journal} {Nature Nanotech}\ }\textbf {\bibinfo {volume} {12}},\ \bibinfo
  {pages} {1026} (\bibinfo {year} {2017})}\BibitemShut {NoStop}%
\bibitem [{\citenamefont {Duan}\ \emph {et~al.}(2001)\citenamefont {Duan},
  \citenamefont {Lukin}, \citenamefont {Cirac},\ and\ \citenamefont
  {Zoller}}]{Duan01n}%
  \BibitemOpen
  \bibfield  {author} {\bibinfo {author} {\bibfnamefont {L.-M.}\ \bibnamefont
  {Duan}}, \bibinfo {author} {\bibfnamefont {M.~D.}\ \bibnamefont {Lukin}},
  \bibinfo {author} {\bibfnamefont {J.~I.}\ \bibnamefont {Cirac}},\ and\
  \bibinfo {author} {\bibfnamefont {P.}~\bibnamefont {Zoller}},\ }\href
  {https://doi.org/10.1038/35106500} {\bibfield  {journal} {\bibinfo  {journal}
  {Nature}\ }\textbf {\bibinfo {volume} {414}},\ \bibinfo {pages} {413}
  (\bibinfo {year} {2001})}\BibitemShut {NoStop}%
\bibitem [{\citenamefont {Reiserer}\ and\ \citenamefont
  {Rempe}(2015)}]{Reiserer15rmp}%
  \BibitemOpen
  \bibfield  {author} {\bibinfo {author} {\bibfnamefont {A.}~\bibnamefont
  {Reiserer}}\ and\ \bibinfo {author} {\bibfnamefont {G.}~\bibnamefont
  {Rempe}},\ }\href {https://doi.org/10.1103/RevModPhys.87.1379} {\bibfield
  {journal} {\bibinfo  {journal} {Rev. Mod. Phys.}\ }\textbf {\bibinfo {volume}
  {87}},\ \bibinfo {pages} {1379} (\bibinfo {year} {2015})}\BibitemShut
  {NoStop}%
\bibitem [{\citenamefont {Hermans}\ \emph {et~al.}(2022)\citenamefont
  {Hermans}, \citenamefont {Pompili}, \citenamefont {Beukers}, \citenamefont
  {Baier}, \citenamefont {Borregaard},\ and\ \citenamefont
  {Hanson}}]{Hermans22n}%
  \BibitemOpen
  \bibfield  {author} {\bibinfo {author} {\bibfnamefont {S.~L.~N.}\
  \bibnamefont {Hermans}}, \bibinfo {author} {\bibfnamefont {M.}~\bibnamefont
  {Pompili}}, \bibinfo {author} {\bibfnamefont {H.~K.~C.}\ \bibnamefont
  {Beukers}}, \bibinfo {author} {\bibfnamefont {S.}~\bibnamefont {Baier}},
  \bibinfo {author} {\bibfnamefont {J.}~\bibnamefont {Borregaard}},\ and\
  \bibinfo {author} {\bibfnamefont {R.}~\bibnamefont {Hanson}},\ }\href
  {https://doi.org/10.1038/s41586-022-04697-y} {\bibfield  {journal} {\bibinfo
  {journal} {Nature}\ }\textbf {\bibinfo {volume} {605}},\ \bibinfo {pages}
  {663} (\bibinfo {year} {2022})}\BibitemShut {NoStop}%
\bibitem [{\citenamefont {Nadlinger}\ \emph {et~al.}(2022)\citenamefont
  {Nadlinger}, \citenamefont {Drmota}, \citenamefont {Nichol}, \citenamefont
  {Araneda}, \citenamefont {Main}, \citenamefont {Srinivas}, \citenamefont
  {Lucas}, \citenamefont {Ballance}, \citenamefont {Ivanov}, \citenamefont
  {Tan}, \citenamefont {Sekatski}, \citenamefont {Urbanke}, \citenamefont
  {Renner}, \citenamefont {Sangouard},\ and\ \citenamefont
  {Bancal}}]{Nadlinger22n}%
  \BibitemOpen
  \bibfield  {author} {\bibinfo {author} {\bibfnamefont {D.~P.}\ \bibnamefont
  {Nadlinger}}, \bibinfo {author} {\bibfnamefont {P.}~\bibnamefont {Drmota}},
  \bibinfo {author} {\bibfnamefont {B.~C.}\ \bibnamefont {Nichol}}, \bibinfo
  {author} {\bibfnamefont {G.}~\bibnamefont {Araneda}}, \bibinfo {author}
  {\bibfnamefont {D.}~\bibnamefont {Main}}, \bibinfo {author} {\bibfnamefont
  {R.}~\bibnamefont {Srinivas}}, \bibinfo {author} {\bibfnamefont {D.~M.}\
  \bibnamefont {Lucas}}, \bibinfo {author} {\bibfnamefont {C.~J.}\ \bibnamefont
  {Ballance}}, \bibinfo {author} {\bibfnamefont {K.}~\bibnamefont {Ivanov}},
  \bibinfo {author} {\bibfnamefont {E.~Y.-Z.}\ \bibnamefont {Tan}}, \bibinfo
  {author} {\bibfnamefont {P.}~\bibnamefont {Sekatski}}, \bibinfo {author}
  {\bibfnamefont {R.~L.}\ \bibnamefont {Urbanke}}, \bibinfo {author}
  {\bibfnamefont {R.}~\bibnamefont {Renner}}, \bibinfo {author} {\bibfnamefont
  {N.}~\bibnamefont {Sangouard}},\ and\ \bibinfo {author} {\bibfnamefont
  {J.-D.}\ \bibnamefont {Bancal}},\ }\href
  {https://doi.org/10.1038/s41586-022-04941-5} {\bibfield  {journal} {\bibinfo
  {journal} {Nature}\ }\textbf {\bibinfo {volume} {607}},\ \bibinfo {pages}
  {682} (\bibinfo {year} {2022})}\BibitemShut {NoStop}%
\bibitem [{\citenamefont {Takahashi}\ \emph {et~al.}(2020)\citenamefont
  {Takahashi}, \citenamefont {Kassa}, \citenamefont {Christoforou},\ and\
  \citenamefont {Keller}}]{Takahashi20prl}%
  \BibitemOpen
  \bibfield  {author} {\bibinfo {author} {\bibfnamefont {H.}~\bibnamefont
  {Takahashi}}, \bibinfo {author} {\bibfnamefont {E.}~\bibnamefont {Kassa}},
  \bibinfo {author} {\bibfnamefont {C.}~\bibnamefont {Christoforou}},\ and\
  \bibinfo {author} {\bibfnamefont {M.}~\bibnamefont {Keller}},\ }\href
  {https://doi.org/10.1103/PhysRevLett.124.013602} {\bibfield  {journal}
  {\bibinfo  {journal} {Phys. Rev. Lett.}\ }\textbf {\bibinfo {volume} {124}},\
  \bibinfo {pages} {013602} (\bibinfo {year} {2020})}\BibitemShut {NoStop}%
\bibitem [{\citenamefont {Brekenfeld}\ \emph {et~al.}(2020)\citenamefont
  {Brekenfeld}, \citenamefont {Niemietz}, \citenamefont {Christesen},\ and\
  \citenamefont {Rempe}}]{Brekenfeld20np}%
  \BibitemOpen
  \bibfield  {author} {\bibinfo {author} {\bibfnamefont {M.}~\bibnamefont
  {Brekenfeld}}, \bibinfo {author} {\bibfnamefont {D.}~\bibnamefont
  {Niemietz}}, \bibinfo {author} {\bibfnamefont {J.~D.}\ \bibnamefont
  {Christesen}},\ and\ \bibinfo {author} {\bibfnamefont {G.}~\bibnamefont
  {Rempe}},\ }\href {https://doi.org/10.1038/s41567-020-0855-3} {\bibfield
  {journal} {\bibinfo  {journal} {Nature Physics}\ }\textbf {\bibinfo {volume}
  {16}},\ \bibinfo {pages} {647} (\bibinfo {year} {2020})}\BibitemShut
  {NoStop}%
\bibitem [{\citenamefont {Goto}\ \emph {et~al.}(2019)\citenamefont {Goto},
  \citenamefont {Mizukami}, \citenamefont {Tokunaga},\ and\ \citenamefont
  {Aoki}}]{goto-aoki2019praa}%
  \BibitemOpen
  \bibfield  {author} {\bibinfo {author} {\bibfnamefont {H.}~\bibnamefont
  {Goto}}, \bibinfo {author} {\bibfnamefont {S.}~\bibnamefont {Mizukami}},
  \bibinfo {author} {\bibfnamefont {Y.}~\bibnamefont {Tokunaga}},\ and\
  \bibinfo {author} {\bibfnamefont {T.}~\bibnamefont {Aoki}},\ }\href
  {https://doi.org/10/ggz5gg} {\bibfield  {journal} {\bibinfo  {journal} {Phys.
  Rev. A}\ }\textbf {\bibinfo {volume} {99}},\ \bibinfo {pages} {053843}
  (\bibinfo {year} {2019})}\BibitemShut {NoStop}%
\bibitem [{\citenamefont {Gao}\ \emph {et~al.}(2023)\citenamefont {Gao},
  \citenamefont {Blackmore}, \citenamefont {Hughes}, \citenamefont {Doherty},\
  and\ \citenamefont {Goodwin}}]{Gao23pra}%
  \BibitemOpen
  \bibfield  {author} {\bibinfo {author} {\bibfnamefont {S.}~\bibnamefont
  {Gao}}, \bibinfo {author} {\bibfnamefont {J.~A.}\ \bibnamefont {Blackmore}},
  \bibinfo {author} {\bibfnamefont {W.~J.}\ \bibnamefont {Hughes}}, \bibinfo
  {author} {\bibfnamefont {T.~H.}\ \bibnamefont {Doherty}},\ and\ \bibinfo
  {author} {\bibfnamefont {J.~F.}\ \bibnamefont {Goodwin}},\ }\href
  {https://doi.org/10.1103/PhysRevApplied.19.014033} {\bibfield  {journal}
  {\bibinfo  {journal} {Phys. Rev. Appl.}\ }\textbf {\bibinfo {volume} {19}},\
  \bibinfo {pages} {014033} (\bibinfo {year} {2023})}\BibitemShut {NoStop}%
\bibitem [{\citenamefont {Hughes}\ \emph {et~al.}(2023)\citenamefont {Hughes},
  \citenamefont {Doherty}, \citenamefont {Blackmore}, \citenamefont {Horak},\
  and\ \citenamefont {Goodwin}}]{Hughes23oe}%
  \BibitemOpen
  \bibfield  {author} {\bibinfo {author} {\bibfnamefont {W.~J.}\ \bibnamefont
  {Hughes}}, \bibinfo {author} {\bibfnamefont {T.~H.}\ \bibnamefont {Doherty}},
  \bibinfo {author} {\bibfnamefont {J.~A.}\ \bibnamefont {Blackmore}}, \bibinfo
  {author} {\bibfnamefont {P.}~\bibnamefont {Horak}},\ and\ \bibinfo {author}
  {\bibfnamefont {J.~F.}\ \bibnamefont {Goodwin}},\ }\href
  {https://doi.org/10.1364/OE.496981} {\bibfield  {journal} {\bibinfo
  {journal} {Opt. Express}\ }\textbf {\bibinfo {volume} {31}},\ \bibinfo
  {pages} {32619} (\bibinfo {year} {2023})}\BibitemShut {NoStop}%
\bibitem [{\citenamefont {Hunger}\ \emph {et~al.}(2010)\citenamefont {Hunger},
  \citenamefont {Steinmetz}, \citenamefont {Colombe}, \citenamefont {Deutsch},
  \citenamefont {H{\"a}nsch},\ and\ \citenamefont {Reichel}}]{Hunger10njp}%
  \BibitemOpen
  \bibfield  {author} {\bibinfo {author} {\bibfnamefont {D.}~\bibnamefont
  {Hunger}}, \bibinfo {author} {\bibfnamefont {T.}~\bibnamefont {Steinmetz}},
  \bibinfo {author} {\bibfnamefont {Y.}~\bibnamefont {Colombe}}, \bibinfo
  {author} {\bibfnamefont {C.}~\bibnamefont {Deutsch}}, \bibinfo {author}
  {\bibfnamefont {T.~W.}\ \bibnamefont {H{\"a}nsch}},\ and\ \bibinfo {author}
  {\bibfnamefont {J.}~\bibnamefont {Reichel}},\ }\href
  {https://doi.org/10.1088/1367-2630/12/6/065038} {\bibfield  {journal}
  {\bibinfo  {journal} {New J. Phys.}\ }\textbf {\bibinfo {volume} {12}},\
  \bibinfo {pages} {065038} (\bibinfo {year} {2010})}\BibitemShut {NoStop}%
\bibitem [{\citenamefont {Rempe}\ \emph {et~al.}(1992)\citenamefont {Rempe},
  \citenamefont {Thompson}, \citenamefont {Kimble},\ and\ \citenamefont
  {Lalezari}}]{rempe-lalezari1992ola}%
  \BibitemOpen
  \bibfield  {author} {\bibinfo {author} {\bibfnamefont {G.}~\bibnamefont
  {Rempe}}, \bibinfo {author} {\bibfnamefont {R.~J.}\ \bibnamefont {Thompson}},
  \bibinfo {author} {\bibfnamefont {H.~J.}\ \bibnamefont {Kimble}},\ and\
  \bibinfo {author} {\bibfnamefont {R.}~\bibnamefont {Lalezari}},\ }\href
  {https://doi.org/10.1364/OL.17.000363} {\bibfield  {journal} {\bibinfo
  {journal} {OPTICS LETTERS}\ }\textbf {\bibinfo {volume} {17}},\ \bibinfo
  {pages} {3} (\bibinfo {year} {1992})}\BibitemShut {NoStop}%
\bibitem [{\citenamefont {Jin}\ \emph {et~al.}(2022)\citenamefont {Jin},
  \citenamefont {McLemore}, \citenamefont {Mason}, \citenamefont {Hendrie},
  \citenamefont {Luo}, \citenamefont {Kelleher}, \citenamefont {Kharel},
  \citenamefont {Quinlan}, \citenamefont {Diddams},\ and\ \citenamefont
  {Rakich}}]{jin-rakich2022oa}%
  \BibitemOpen
  \bibfield  {author} {\bibinfo {author} {\bibfnamefont {N.}~\bibnamefont
  {Jin}}, \bibinfo {author} {\bibfnamefont {C.~A.}\ \bibnamefont {McLemore}},
  \bibinfo {author} {\bibfnamefont {D.}~\bibnamefont {Mason}}, \bibinfo
  {author} {\bibfnamefont {J.~P.}\ \bibnamefont {Hendrie}}, \bibinfo {author}
  {\bibfnamefont {Y.}~\bibnamefont {Luo}}, \bibinfo {author} {\bibfnamefont
  {M.~L.}\ \bibnamefont {Kelleher}}, \bibinfo {author} {\bibfnamefont
  {P.}~\bibnamefont {Kharel}}, \bibinfo {author} {\bibfnamefont
  {F.}~\bibnamefont {Quinlan}}, \bibinfo {author} {\bibfnamefont {S.~A.}\
  \bibnamefont {Diddams}},\ and\ \bibinfo {author} {\bibfnamefont {P.~T.}\
  \bibnamefont {Rakich}},\ }\href {https://doi.org/10.1364/OPTICA.467440}
  {\bibfield  {journal} {\bibinfo  {journal} {Optica}\ }\textbf {\bibinfo
  {volume} {9}},\ \bibinfo {pages} {965} (\bibinfo {year} {2022})}\BibitemShut
  {NoStop}%
\bibitem [{\citenamefont {Gallego}\ \emph {et~al.}(2016)\citenamefont
  {Gallego}, \citenamefont {Ghosh}, \citenamefont {Alavi}, \citenamefont {Alt},
  \citenamefont {{Martinez-Dorantes}}, \citenamefont {Meschede},\ and\
  \citenamefont {Ratschbacher}}]{Gallego16apb}%
  \BibitemOpen
  \bibfield  {author} {\bibinfo {author} {\bibfnamefont {J.}~\bibnamefont
  {Gallego}}, \bibinfo {author} {\bibfnamefont {S.}~\bibnamefont {Ghosh}},
  \bibinfo {author} {\bibfnamefont {S.~K.}\ \bibnamefont {Alavi}}, \bibinfo
  {author} {\bibfnamefont {W.}~\bibnamefont {Alt}}, \bibinfo {author}
  {\bibfnamefont {M.}~\bibnamefont {{Martinez-Dorantes}}}, \bibinfo {author}
  {\bibfnamefont {D.}~\bibnamefont {Meschede}},\ and\ \bibinfo {author}
  {\bibfnamefont {L.}~\bibnamefont {Ratschbacher}},\ }\href
  {https://doi.org/10.1007/s00340-015-6281-z} {\bibfield  {journal} {\bibinfo
  {journal} {Appl. Phys. B}\ }\textbf {\bibinfo {volume} {122}},\ \bibinfo
  {pages} {47} (\bibinfo {year} {2016})}\BibitemShut {NoStop}%
\bibitem [{\citenamefont {Bick}\ \emph {et~al.}(2016)\citenamefont {Bick},
  \citenamefont {Staarmann}, \citenamefont {Christoph}, \citenamefont
  {Hellmig}, \citenamefont {Heinze}, \citenamefont {Sengstock},\ and\
  \citenamefont {Becker}}]{Bick16rosi}%
  \BibitemOpen
  \bibfield  {author} {\bibinfo {author} {\bibfnamefont {A.}~\bibnamefont
  {Bick}}, \bibinfo {author} {\bibfnamefont {C.}~\bibnamefont {Staarmann}},
  \bibinfo {author} {\bibfnamefont {P.}~\bibnamefont {Christoph}}, \bibinfo
  {author} {\bibfnamefont {O.}~\bibnamefont {Hellmig}}, \bibinfo {author}
  {\bibfnamefont {J.}~\bibnamefont {Heinze}}, \bibinfo {author} {\bibfnamefont
  {K.}~\bibnamefont {Sengstock}},\ and\ \bibinfo {author} {\bibfnamefont
  {C.}~\bibnamefont {Becker}},\ }\href {https://doi.org/10.1063/1.4939046}
  {\bibfield  {journal} {\bibinfo  {journal} {Review of Scientific
  Instruments}\ }\textbf {\bibinfo {volume} {87}},\ \bibinfo {pages} {013102}
  (\bibinfo {year} {2016})}\BibitemShut {NoStop}%
\bibitem [{\citenamefont {Pfeifer}\ \emph {et~al.}(2022)\citenamefont
  {Pfeifer}, \citenamefont {Ratschbacher}, \citenamefont {Gallego},
  \citenamefont {Saavedra}, \citenamefont {Fa{\ss}bender}, \citenamefont {{von
  Haaren}}, \citenamefont {Alt}, \citenamefont {Hofferberth}, \citenamefont
  {K{\"o}hl}, \citenamefont {Linden},\ and\ \citenamefont
  {Meschede}}]{Pfeifer22apb}%
  \BibitemOpen
  \bibfield  {author} {\bibinfo {author} {\bibfnamefont {H.}~\bibnamefont
  {Pfeifer}}, \bibinfo {author} {\bibfnamefont {L.}~\bibnamefont
  {Ratschbacher}}, \bibinfo {author} {\bibfnamefont {J.}~\bibnamefont
  {Gallego}}, \bibinfo {author} {\bibfnamefont {C.}~\bibnamefont {Saavedra}},
  \bibinfo {author} {\bibfnamefont {A.}~\bibnamefont {Fa{\ss}bender}}, \bibinfo
  {author} {\bibfnamefont {A.}~\bibnamefont {{von Haaren}}}, \bibinfo {author}
  {\bibfnamefont {W.}~\bibnamefont {Alt}}, \bibinfo {author} {\bibfnamefont
  {S.}~\bibnamefont {Hofferberth}}, \bibinfo {author} {\bibfnamefont
  {M.}~\bibnamefont {K{\"o}hl}}, \bibinfo {author} {\bibfnamefont
  {S.}~\bibnamefont {Linden}},\ and\ \bibinfo {author} {\bibfnamefont
  {D.}~\bibnamefont {Meschede}},\ }\href
  {https://doi.org/10.1007/s00340-022-07752-8} {\bibfield  {journal} {\bibinfo
  {journal} {Appl. Phys. B}\ }\textbf {\bibinfo {volume} {128}},\ \bibinfo
  {pages} {29} (\bibinfo {year} {2022})}\BibitemShut {NoStop}%
\bibitem [{\citenamefont {Gulati}\ \emph {et~al.}(2017)\citenamefont {Gulati},
  \citenamefont {Takahashi}, \citenamefont {Podoliak}, \citenamefont {Horak},\
  and\ \citenamefont {Keller}}]{Gulati17sr}%
  \BibitemOpen
  \bibfield  {author} {\bibinfo {author} {\bibfnamefont {G.~K.}\ \bibnamefont
  {Gulati}}, \bibinfo {author} {\bibfnamefont {H.}~\bibnamefont {Takahashi}},
  \bibinfo {author} {\bibfnamefont {N.}~\bibnamefont {Podoliak}}, \bibinfo
  {author} {\bibfnamefont {P.}~\bibnamefont {Horak}},\ and\ \bibinfo {author}
  {\bibfnamefont {M.}~\bibnamefont {Keller}},\ }\href
  {https://doi.org/10.1038/s41598-017-05729-8} {\bibfield  {journal} {\bibinfo
  {journal} {Sci Rep}\ }\textbf {\bibinfo {volume} {7}},\ \bibinfo {pages}
  {5556} (\bibinfo {year} {2017})}\BibitemShut {NoStop}%
\bibitem [{\citenamefont {Chen}\ \emph {et~al.}(2016)\citenamefont {Chen},
  \citenamefont {Taylor},\ and\ \citenamefont {Yu}}]{Chen16rpp}%
  \BibitemOpen
  \bibfield  {author} {\bibinfo {author} {\bibfnamefont {H.-T.}\ \bibnamefont
  {Chen}}, \bibinfo {author} {\bibfnamefont {A.~J.}\ \bibnamefont {Taylor}},\
  and\ \bibinfo {author} {\bibfnamefont {N.}~\bibnamefont {Yu}},\ }\href
  {https://doi.org/10.1088/0034-4885/79/7/076401} {\bibfield  {journal}
  {\bibinfo  {journal} {Rep. Prog. Phys.}\ }\textbf {\bibinfo {volume} {79}},\
  \bibinfo {pages} {076401} (\bibinfo {year} {2016})}\BibitemShut {NoStop}%
\bibitem [{\citenamefont {Phan}\ \emph {et~al.}(2019)\citenamefont {Phan},
  \citenamefont {Sell}, \citenamefont {Wang}, \citenamefont {Doshay},
  \citenamefont {Edee}, \citenamefont {Yang},\ and\ \citenamefont
  {Fan}}]{phan-fan2019lsaa}%
  \BibitemOpen
  \bibfield  {author} {\bibinfo {author} {\bibfnamefont {T.}~\bibnamefont
  {Phan}}, \bibinfo {author} {\bibfnamefont {D.}~\bibnamefont {Sell}}, \bibinfo
  {author} {\bibfnamefont {E.~W.}\ \bibnamefont {Wang}}, \bibinfo {author}
  {\bibfnamefont {S.}~\bibnamefont {Doshay}}, \bibinfo {author} {\bibfnamefont
  {K.}~\bibnamefont {Edee}}, \bibinfo {author} {\bibfnamefont {J.}~\bibnamefont
  {Yang}},\ and\ \bibinfo {author} {\bibfnamefont {J.~A.}\ \bibnamefont
  {Fan}},\ }\href {https://doi.org/10.1038/s41377-019-0159-5} {\bibfield
  {journal} {\bibinfo  {journal} {Light Sci Appl}\ }\textbf {\bibinfo {volume}
  {8}},\ \bibinfo {pages} {48} (\bibinfo {year} {2019})}\BibitemShut {NoStop}%
\bibitem [{\citenamefont {Hsu}\ \emph {et~al.}(2022)\citenamefont {Hsu},
  \citenamefont {Zhu}, \citenamefont {Thiele}, \citenamefont {Brown},
  \citenamefont {Papp}, \citenamefont {Agrawal},\ and\ \citenamefont
  {Regal}}]{Hsu22pq}%
  \BibitemOpen
  \bibfield  {author} {\bibinfo {author} {\bibfnamefont {T.-W.}\ \bibnamefont
  {Hsu}}, \bibinfo {author} {\bibfnamefont {W.}~\bibnamefont {Zhu}}, \bibinfo
  {author} {\bibfnamefont {T.}~\bibnamefont {Thiele}}, \bibinfo {author}
  {\bibfnamefont {M.~O.}\ \bibnamefont {Brown}}, \bibinfo {author}
  {\bibfnamefont {S.~B.}\ \bibnamefont {Papp}}, \bibinfo {author}
  {\bibfnamefont {A.}~\bibnamefont {Agrawal}},\ and\ \bibinfo {author}
  {\bibfnamefont {C.~A.}\ \bibnamefont {Regal}},\ }\href
  {https://doi.org/10.1103/PRXQuantum.3.030316} {\bibfield  {journal} {\bibinfo
   {journal} {PRX Quantum}\ }\textbf {\bibinfo {volume} {3}},\ \bibinfo {pages}
  {030316} (\bibinfo {year} {2022})}\BibitemShut {NoStop}%
\bibitem [{\citenamefont {Huang}\ \emph {et~al.}(2024)\citenamefont {Huang},
  \citenamefont {Zhou}, \citenamefont {Li}, \citenamefont {Xu}, \citenamefont
  {Wang},\ and\ \citenamefont {Zhan}}]{Huang24oe}%
  \BibitemOpen
  \bibfield  {author} {\bibinfo {author} {\bibfnamefont {R.}~\bibnamefont
  {Huang}}, \bibinfo {author} {\bibfnamefont {F.}~\bibnamefont {Zhou}},
  \bibinfo {author} {\bibfnamefont {X.}~\bibnamefont {Li}}, \bibinfo {author}
  {\bibfnamefont {P.}~\bibnamefont {Xu}}, \bibinfo {author} {\bibfnamefont
  {Y.}~\bibnamefont {Wang}},\ and\ \bibinfo {author} {\bibfnamefont
  {M.}~\bibnamefont {Zhan}},\ }\href {https://doi.org/10.1364/OE.525454}
  {\bibfield  {journal} {\bibinfo  {journal} {Opt. Express}\ }\textbf {\bibinfo
  {volume} {32}},\ \bibinfo {pages} {21293} (\bibinfo {year}
  {2024})}\BibitemShut {NoStop}%
\bibitem [{\citenamefont {Chen}\ \emph {et~al.}(2024)\citenamefont {Chen},
  \citenamefont {Zhao}, \citenamefont {Wang}, \citenamefont {Li}, \citenamefont
  {Zhang}, \citenamefont {Chen}, \citenamefont {Zhang}, \citenamefont {Xu},
  \citenamefont {Liu}, \citenamefont {Dong}, \citenamefont {Guo}, \citenamefont
  {Huang},\ and\ \citenamefont {Zou}}]{Chen24lpr}%
  \BibitemOpen
  \bibfield  {author} {\bibinfo {author} {\bibfnamefont {G.-J.}\ \bibnamefont
  {Chen}}, \bibinfo {author} {\bibfnamefont {D.}~\bibnamefont {Zhao}}, \bibinfo
  {author} {\bibfnamefont {Z.-B.}\ \bibnamefont {Wang}}, \bibinfo {author}
  {\bibfnamefont {Z.}~\bibnamefont {Li}}, \bibinfo {author} {\bibfnamefont
  {J.-Z.}\ \bibnamefont {Zhang}}, \bibinfo {author} {\bibfnamefont
  {L.}~\bibnamefont {Chen}}, \bibinfo {author} {\bibfnamefont {Y.-L.}\
  \bibnamefont {Zhang}}, \bibinfo {author} {\bibfnamefont {X.-B.}\ \bibnamefont
  {Xu}}, \bibinfo {author} {\bibfnamefont {A.-P.}\ \bibnamefont {Liu}},
  \bibinfo {author} {\bibfnamefont {C.-H.}\ \bibnamefont {Dong}}, \bibinfo
  {author} {\bibfnamefont {G.-C.}\ \bibnamefont {Guo}}, \bibinfo {author}
  {\bibfnamefont {K.}~\bibnamefont {Huang}},\ and\ \bibinfo {author}
  {\bibfnamefont {C.-L.}\ \bibnamefont {Zou}},\ }\href
  {https://doi.org/10.1002/lpor.202401595} {\bibfield  {journal} {\bibinfo
  {journal} {Laser \& Photonics Reviews}\ }\textbf {\bibinfo {volume} {n/a}},\
  \bibinfo {pages} {2401595} (\bibinfo {year} {2024})}\BibitemShut {NoStop}%
\bibitem [{\citenamefont {Lim}\ \emph {et~al.}(2024)\citenamefont {Lim},
  \citenamefont {Froech}, \citenamefont {Choi}, \citenamefont {Majumdar},\ and\
  \citenamefont {Mouradian}}]{Lim24q2ce2pq}%
  \BibitemOpen
  \bibfield  {author} {\bibinfo {author} {\bibfnamefont {H.}~\bibnamefont
  {Lim}}, \bibinfo {author} {\bibfnamefont {J.}~\bibnamefont {Froech}},
  \bibinfo {author} {\bibfnamefont {M.}~\bibnamefont {Choi}}, \bibinfo {author}
  {\bibfnamefont {A.}~\bibnamefont {Majumdar}},\ and\ \bibinfo {author}
  {\bibfnamefont {S.}~\bibnamefont {Mouradian}},\ }in\ \href
  {https://doi.org/10.1364/QUANTUM.2024.QM3A.3} {\emph {\bibinfo {booktitle}
  {Quantum 2.0 {{Conference}} and {{Exhibition}} (2024), Paper {{QM3A}}.3}}}\
  (\bibinfo  {publisher} {Optica Publishing Group},\ \bibinfo {year} {2024})\
  p.\ \bibinfo {pages} {QM3A.3}\BibitemShut {NoStop}%
\bibitem [{\citenamefont {Ren}\ \emph {et~al.}(2019)\citenamefont {Ren},
  \citenamefont {Briere}, \citenamefont {Fang}, \citenamefont {Ni},
  \citenamefont {Sawant}, \citenamefont {H{\'e}ron}, \citenamefont {Chenot},
  \citenamefont {V{\'e}zian}, \citenamefont {Damilano}, \citenamefont
  {Br{\"a}ndli}, \citenamefont {Maier},\ and\ \citenamefont
  {Genevet}}]{Ren19nc}%
  \BibitemOpen
  \bibfield  {author} {\bibinfo {author} {\bibfnamefont {H.}~\bibnamefont
  {Ren}}, \bibinfo {author} {\bibfnamefont {G.}~\bibnamefont {Briere}},
  \bibinfo {author} {\bibfnamefont {X.}~\bibnamefont {Fang}}, \bibinfo {author}
  {\bibfnamefont {P.}~\bibnamefont {Ni}}, \bibinfo {author} {\bibfnamefont
  {R.}~\bibnamefont {Sawant}}, \bibinfo {author} {\bibfnamefont
  {S.}~\bibnamefont {H{\'e}ron}}, \bibinfo {author} {\bibfnamefont
  {S.}~\bibnamefont {Chenot}}, \bibinfo {author} {\bibfnamefont
  {S.}~\bibnamefont {V{\'e}zian}}, \bibinfo {author} {\bibfnamefont
  {B.}~\bibnamefont {Damilano}}, \bibinfo {author} {\bibfnamefont
  {V.}~\bibnamefont {Br{\"a}ndli}}, \bibinfo {author} {\bibfnamefont {S.~A.}\
  \bibnamefont {Maier}},\ and\ \bibinfo {author} {\bibfnamefont
  {P.}~\bibnamefont {Genevet}},\ }\href
  {https://doi.org/10.1038/s41467-019-11030-1} {\bibfield  {journal} {\bibinfo
  {journal} {Nat Commun}\ }\textbf {\bibinfo {volume} {10}},\ \bibinfo {pages}
  {2986} (\bibinfo {year} {2019})}\BibitemShut {NoStop}%
\bibitem [{\citenamefont {Zhou}\ \emph {et~al.}(2021)\citenamefont {Zhou},
  \citenamefont {Liu}, \citenamefont {Zhu}, \citenamefont {Chen}, \citenamefont
  {Ren}, \citenamefont {Lezec}, \citenamefont {Lu}, \citenamefont {Agrawal},\
  and\ \citenamefont {Xu}}]{Zhou21lpr}%
  \BibitemOpen
  \bibfield  {author} {\bibinfo {author} {\bibfnamefont {Q.}~\bibnamefont
  {Zhou}}, \bibinfo {author} {\bibfnamefont {M.}~\bibnamefont {Liu}}, \bibinfo
  {author} {\bibfnamefont {W.}~\bibnamefont {Zhu}}, \bibinfo {author}
  {\bibfnamefont {L.}~\bibnamefont {Chen}}, \bibinfo {author} {\bibfnamefont
  {Y.}~\bibnamefont {Ren}}, \bibinfo {author} {\bibfnamefont {H.~J.}\
  \bibnamefont {Lezec}}, \bibinfo {author} {\bibfnamefont {Y.}~\bibnamefont
  {Lu}}, \bibinfo {author} {\bibfnamefont {A.}~\bibnamefont {Agrawal}},\ and\
  \bibinfo {author} {\bibfnamefont {T.}~\bibnamefont {Xu}},\ }\href
  {https://doi.org/10.1002/lpor.202100390} {\bibfield  {journal} {\bibinfo
  {journal} {Laser \& Photonics Reviews}\ }\textbf {\bibinfo {volume} {15}},\
  \bibinfo {pages} {2100390} (\bibinfo {year} {2021})}\BibitemShut {NoStop}%
\bibitem [{\citenamefont {Ossiander}\ \emph {et~al.}(2023)\citenamefont
  {Ossiander}, \citenamefont {Meretska}, \citenamefont {Rourke}, \citenamefont
  {Sp{\"a}gele}, \citenamefont {Yin}, \citenamefont {{Benea-Chelmus}},\ and\
  \citenamefont {Capasso}}]{Ossiander23nc}%
  \BibitemOpen
  \bibfield  {author} {\bibinfo {author} {\bibfnamefont {M.}~\bibnamefont
  {Ossiander}}, \bibinfo {author} {\bibfnamefont {M.~L.}\ \bibnamefont
  {Meretska}}, \bibinfo {author} {\bibfnamefont {S.}~\bibnamefont {Rourke}},
  \bibinfo {author} {\bibfnamefont {C.}~\bibnamefont {Sp{\"a}gele}}, \bibinfo
  {author} {\bibfnamefont {X.}~\bibnamefont {Yin}}, \bibinfo {author}
  {\bibfnamefont {I.-C.}\ \bibnamefont {{Benea-Chelmus}}},\ and\ \bibinfo
  {author} {\bibfnamefont {F.}~\bibnamefont {Capasso}},\ }\href
  {https://doi.org/10.1038/s41467-023-36873-7} {\bibfield  {journal} {\bibinfo
  {journal} {Nat Commun}\ }\textbf {\bibinfo {volume} {14}},\ \bibinfo {pages}
  {1114} (\bibinfo {year} {2023})}\BibitemShut {NoStop}%
\bibitem [{\citenamefont {Fontana}\ \emph {et~al.}(2021)\citenamefont
  {Fontana}, \citenamefont {Zifkin}, \citenamefont {Janitz}, \citenamefont
  {Rodr{\'i}guez~Rosenblueth},\ and\ \citenamefont
  {Childress}}]{Fontana21rosi}%
  \BibitemOpen
  \bibfield  {author} {\bibinfo {author} {\bibfnamefont {Y.}~\bibnamefont
  {Fontana}}, \bibinfo {author} {\bibfnamefont {R.}~\bibnamefont {Zifkin}},
  \bibinfo {author} {\bibfnamefont {E.}~\bibnamefont {Janitz}}, \bibinfo
  {author} {\bibfnamefont {C.~D.}\ \bibnamefont {Rodr{\'i}guez~Rosenblueth}},\
  and\ \bibinfo {author} {\bibfnamefont {L.}~\bibnamefont {Childress}},\ }\href
  {https://doi.org/10.1063/5.0049520} {\bibfield  {journal} {\bibinfo
  {journal} {Review of Scientific Instruments}\ }\textbf {\bibinfo {volume}
  {92}},\ \bibinfo {pages} {053906} (\bibinfo {year} {2021})}\BibitemShut
  {NoStop}%
\bibitem [{\citenamefont {Wipfli}\ \emph {et~al.}(2023)\citenamefont {Wipfli},
  \citenamefont {Passagem}, \citenamefont {Fischer}, \citenamefont {Grau},\
  and\ \citenamefont {Home}}]{Wipfli23rosi}%
  \BibitemOpen
  \bibfield  {author} {\bibinfo {author} {\bibfnamefont {O.}~\bibnamefont
  {Wipfli}}, \bibinfo {author} {\bibfnamefont {H.~F.}\ \bibnamefont
  {Passagem}}, \bibinfo {author} {\bibfnamefont {C.}~\bibnamefont {Fischer}},
  \bibinfo {author} {\bibfnamefont {M.}~\bibnamefont {Grau}},\ and\ \bibinfo
  {author} {\bibfnamefont {J.~P.}\ \bibnamefont {Home}},\ }\href
  {https://doi.org/10.1063/5.0155418} {\bibfield  {journal} {\bibinfo
  {journal} {Review of Scientific Instruments}\ }\textbf {\bibinfo {volume}
  {94}},\ \bibinfo {pages} {083204} (\bibinfo {year} {2023})}\BibitemShut
  {NoStop}%
\bibitem [{\citenamefont {Kumar}\ \emph {et~al.}(2023)\citenamefont {Kumar},
  \citenamefont {Suleymanzade}, \citenamefont {Stone}, \citenamefont {Taneja},
  \citenamefont {Anferov}, \citenamefont {Schuster},\ and\ \citenamefont
  {Simon}}]{Kumar23n}%
  \BibitemOpen
  \bibfield  {author} {\bibinfo {author} {\bibfnamefont {A.}~\bibnamefont
  {Kumar}}, \bibinfo {author} {\bibfnamefont {A.}~\bibnamefont {Suleymanzade}},
  \bibinfo {author} {\bibfnamefont {M.}~\bibnamefont {Stone}}, \bibinfo
  {author} {\bibfnamefont {L.}~\bibnamefont {Taneja}}, \bibinfo {author}
  {\bibfnamefont {A.}~\bibnamefont {Anferov}}, \bibinfo {author} {\bibfnamefont
  {D.~I.}\ \bibnamefont {Schuster}},\ and\ \bibinfo {author} {\bibfnamefont
  {J.}~\bibnamefont {Simon}},\ }\href
  {https://doi.org/10.1038/s41586-023-05740-2} {\bibfield  {journal} {\bibinfo
  {journal} {Nature}\ }\textbf {\bibinfo {volume} {615}},\ \bibinfo {pages}
  {614} (\bibinfo {year} {2023})}\BibitemShut {NoStop}%
\bibitem [{\citenamefont {Nguyen}\ \emph {et~al.}(2018)\citenamefont {Nguyen},
  \citenamefont {Utama}, \citenamefont {Lewty},\ and\ \citenamefont
  {Kurtsiefer}}]{Nguyen18pra}%
  \BibitemOpen
  \bibfield  {author} {\bibinfo {author} {\bibfnamefont {C.~H.}\ \bibnamefont
  {Nguyen}}, \bibinfo {author} {\bibfnamefont {A.~N.}\ \bibnamefont {Utama}},
  \bibinfo {author} {\bibfnamefont {N.}~\bibnamefont {Lewty}},\ and\ \bibinfo
  {author} {\bibfnamefont {C.}~\bibnamefont {Kurtsiefer}},\ }\href
  {https://doi.org/10.1103/PhysRevA.98.063833} {\bibfield  {journal} {\bibinfo
  {journal} {Phys. Rev. A}\ }\textbf {\bibinfo {volume} {98}},\ \bibinfo
  {pages} {063833} (\bibinfo {year} {2018})}\BibitemShut {NoStop}%
\bibitem [{\citenamefont {Krutyanskiy}\ \emph {et~al.}(2023)\citenamefont
  {Krutyanskiy}, \citenamefont {Canteri}, \citenamefont {Meraner},
  \citenamefont {Bate}, \citenamefont {Krcmarsky}, \citenamefont {Schupp},
  \citenamefont {Sangouard},\ and\ \citenamefont {Lanyon}}]{Krutyanskiy23prl}%
  \BibitemOpen
  \bibfield  {author} {\bibinfo {author} {\bibfnamefont {V.}~\bibnamefont
  {Krutyanskiy}}, \bibinfo {author} {\bibfnamefont {M.}~\bibnamefont
  {Canteri}}, \bibinfo {author} {\bibfnamefont {M.}~\bibnamefont {Meraner}},
  \bibinfo {author} {\bibfnamefont {J.}~\bibnamefont {Bate}}, \bibinfo {author}
  {\bibfnamefont {V.}~\bibnamefont {Krcmarsky}}, \bibinfo {author}
  {\bibfnamefont {J.}~\bibnamefont {Schupp}}, \bibinfo {author} {\bibfnamefont
  {N.}~\bibnamefont {Sangouard}},\ and\ \bibinfo {author} {\bibfnamefont
  {B.~P.}\ \bibnamefont {Lanyon}},\ }\href
  {https://doi.org/10.1103/PhysRevLett.130.213601} {\bibfield  {journal}
  {\bibinfo  {journal} {Phys. Rev. Lett.}\ }\textbf {\bibinfo {volume} {130}},\
  \bibinfo {pages} {213601} (\bibinfo {year} {2023})}\BibitemShut {NoStop}%
\bibitem [{\citenamefont {Periwal}\ \emph {et~al.}(2021)\citenamefont
  {Periwal}, \citenamefont {Cooper}, \citenamefont {Kunkel}, \citenamefont
  {Wienand}, \citenamefont {Davis},\ and\ \citenamefont
  {{Schleier-Smith}}}]{Periwal21n}%
  \BibitemOpen
  \bibfield  {author} {\bibinfo {author} {\bibfnamefont {A.}~\bibnamefont
  {Periwal}}, \bibinfo {author} {\bibfnamefont {E.~S.}\ \bibnamefont {Cooper}},
  \bibinfo {author} {\bibfnamefont {P.}~\bibnamefont {Kunkel}}, \bibinfo
  {author} {\bibfnamefont {J.~F.}\ \bibnamefont {Wienand}}, \bibinfo {author}
  {\bibfnamefont {E.~J.}\ \bibnamefont {Davis}},\ and\ \bibinfo {author}
  {\bibfnamefont {M.}~\bibnamefont {{Schleier-Smith}}},\ }\href
  {https://doi.org/10.1038/s41586-021-04156-0} {\bibfield  {journal} {\bibinfo
  {journal} {Nature}\ }\textbf {\bibinfo {volume} {600}},\ \bibinfo {pages}
  {630} (\bibinfo {year} {2021})}\BibitemShut {NoStop}%
\bibitem [{\citenamefont {Deist}\ \emph {et~al.}(2022)\citenamefont {Deist},
  \citenamefont {Lu}, \citenamefont {Ho}, \citenamefont {Pasha}, \citenamefont
  {Zeiher}, \citenamefont {Yan},\ and\ \citenamefont
  {{Stamper-Kurn}}}]{Deist22prl}%
  \BibitemOpen
  \bibfield  {author} {\bibinfo {author} {\bibfnamefont {E.}~\bibnamefont
  {Deist}}, \bibinfo {author} {\bibfnamefont {Y.-H.}\ \bibnamefont {Lu}},
  \bibinfo {author} {\bibfnamefont {J.}~\bibnamefont {Ho}}, \bibinfo {author}
  {\bibfnamefont {M.~K.}\ \bibnamefont {Pasha}}, \bibinfo {author}
  {\bibfnamefont {J.}~\bibnamefont {Zeiher}}, \bibinfo {author} {\bibfnamefont
  {Z.}~\bibnamefont {Yan}},\ and\ \bibinfo {author} {\bibfnamefont {D.~M.}\
  \bibnamefont {{Stamper-Kurn}}},\ }\href
  {https://doi.org/10.1103/PhysRevLett.129.203602} {\bibfield  {journal}
  {\bibinfo  {journal} {Phys. Rev. Lett.}\ }\textbf {\bibinfo {volume} {129}},\
  \bibinfo {pages} {203602} (\bibinfo {year} {2022})}\BibitemShut {NoStop}%
\bibitem [{\citenamefont {Shadmany}\ \emph {et~al.}(2025)\citenamefont
  {Shadmany}, \citenamefont {Kumar}, \citenamefont {Soper}, \citenamefont
  {Palm}, \citenamefont {Yin}, \citenamefont {Ando}, \citenamefont {Li},
  \citenamefont {Taneja}, \citenamefont {Jaffe}, \citenamefont {David},\ and\
  \citenamefont {Simon}}]{Shadmany25sa}%
  \BibitemOpen
  \bibfield  {author} {\bibinfo {author} {\bibfnamefont {D.}~\bibnamefont
  {Shadmany}}, \bibinfo {author} {\bibfnamefont {A.}~\bibnamefont {Kumar}},
  \bibinfo {author} {\bibfnamefont {A.}~\bibnamefont {Soper}}, \bibinfo
  {author} {\bibfnamefont {L.}~\bibnamefont {Palm}}, \bibinfo {author}
  {\bibfnamefont {C.}~\bibnamefont {Yin}}, \bibinfo {author} {\bibfnamefont
  {H.}~\bibnamefont {Ando}}, \bibinfo {author} {\bibfnamefont {B.}~\bibnamefont
  {Li}}, \bibinfo {author} {\bibfnamefont {L.}~\bibnamefont {Taneja}}, \bibinfo
  {author} {\bibfnamefont {M.}~\bibnamefont {Jaffe}}, \bibinfo {author}
  {\bibfnamefont {S.}~\bibnamefont {David}},\ and\ \bibinfo {author}
  {\bibfnamefont {J.}~\bibnamefont {Simon}},\ }\href
  {https://doi.org/10.1126/sciadv.ads8171} {\bibfield  {journal} {\bibinfo
  {journal} {Science Advances}\ }\textbf {\bibinfo {volume} {11}},\ \bibinfo
  {pages} {eads8171} (\bibinfo {year} {2025})}\BibitemShut {NoStop}%
\bibitem [{\citenamefont {Corning}(2024)}]{Corning24Web}%
  \BibitemOpen
  \bibfield  {author} {\bibinfo {author} {\bibnamefont {Corning}},\ }\href@noop
  {} {\bibinfo {title} {{{SMF-28 Ultra Optical Fibers}} {\textbar} {{SMF-28
  Ultra}} 200 and 242 {$M$}m {{Single-mode Optical Fiber}} {\textbar}
  {{Corning}}}},\ \bibinfo {howpublished}
  {https://www.corning.com/optical-communications/worldwide/en/home/products/fiber/optical-fiber-products/smf-28-ultra.html}
  (\bibinfo {year} {2024})\BibitemShut {NoStop}%
\bibitem [{\citenamefont {Joyce}\ and\ \citenamefont
  {DeLoach}(1984)}]{Joyce84ao}%
  \BibitemOpen
  \bibfield  {author} {\bibinfo {author} {\bibfnamefont {W.~B.}\ \bibnamefont
  {Joyce}}\ and\ \bibinfo {author} {\bibfnamefont {B.~C.}\ \bibnamefont
  {DeLoach}},\ }\href {https://doi.org/10.1364/AO.23.004187} {\bibfield
  {journal} {\bibinfo  {journal} {Appl. Opt.}\ }\textbf {\bibinfo {volume}
  {23}},\ \bibinfo {pages} {4187} (\bibinfo {year} {1984})}\BibitemShut
  {NoStop}%
\bibitem [{no-()}]{no-endorsement}%
  \BibitemOpen
  \href@noop {} {\bibinfo {title} {Products or companies named here are
  included in the interest of completeness and does not imply endorsement by
  the authors.}}\BibitemShut {Stop}%
\bibitem [{\citenamefont {Egede~Johansen}\ \emph {et~al.}(2024)\citenamefont
  {Egede~Johansen}, \citenamefont {G{\"u}r}, \citenamefont
  {{Mart{\'i}nez-Llin{\'a}s}}, \citenamefont {Fly~Hansen}, \citenamefont
  {Samadi}, \citenamefont {Skak Vestergaard~Larsen}, \citenamefont {Nielsen},
  \citenamefont {Mattinson}, \citenamefont {Schmidlin}, \citenamefont
  {Mortensen},\ and\ \citenamefont {Quaade}}]{EgedeJohansen24cp}%
  \BibitemOpen
  \bibfield  {author} {\bibinfo {author} {\bibfnamefont {V.}~\bibnamefont
  {Egede~Johansen}}, \bibinfo {author} {\bibfnamefont {U.~M.}\ \bibnamefont
  {G{\"u}r}}, \bibinfo {author} {\bibfnamefont {J.}~\bibnamefont
  {{Mart{\'i}nez-Llin{\'a}s}}}, \bibinfo {author} {\bibfnamefont
  {J.}~\bibnamefont {Fly~Hansen}}, \bibinfo {author} {\bibfnamefont
  {A.}~\bibnamefont {Samadi}}, \bibinfo {author} {\bibfnamefont
  {M.}~\bibnamefont {Skak Vestergaard~Larsen}}, \bibinfo {author}
  {\bibfnamefont {T.}~\bibnamefont {Nielsen}}, \bibinfo {author} {\bibfnamefont
  {F.}~\bibnamefont {Mattinson}}, \bibinfo {author} {\bibfnamefont
  {M.}~\bibnamefont {Schmidlin}}, \bibinfo {author} {\bibfnamefont {N.~A.}\
  \bibnamefont {Mortensen}},\ and\ \bibinfo {author} {\bibfnamefont {U.~J.}\
  \bibnamefont {Quaade}},\ }\href {https://doi.org/10.1038/s42005-024-01598-6}
  {\bibfield  {journal} {\bibinfo  {journal} {Commun Phys}\ }\textbf {\bibinfo
  {volume} {7}},\ \bibinfo {pages} {1} (\bibinfo {year} {2024})}\BibitemShut
  {NoStop}%
\bibitem [{\citenamefont {Norland}(2024)}]{Norland24Web}%
  \BibitemOpen
  \bibfield  {author} {\bibinfo {author} {\bibnamefont {Norland}},\ }\href@noop
  {} {\bibinfo {title} {{{NOA}} 61 {\textbar} {{Norland Products}},
  {{Inc}}.}},\ \bibinfo {howpublished}
  {https://norlandproducts.com/product/noa-61/} (\bibinfo {year}
  {2024})\BibitemShut {NoStop}%
\bibitem [{\citenamefont {Masterbond}(2024)}]{Masterbond24Web}%
  \BibitemOpen
  \bibfield  {author} {\bibinfo {author} {\bibnamefont {Masterbond}},\
  }\href@noop {} {\bibinfo {title} {{{EP42HT-3AO Product Information}}
  {\textbar} {{MasterBond}}.com}},\ \bibinfo {howpublished}
  {https://www.masterbond.com/tds/ep42ht-3ao} (\bibinfo {year}
  {2024})\BibitemShut {NoStop}%
\bibitem [{\citenamefont {Adams}\ \emph {et~al.}(2019)\citenamefont {Adams},
  \citenamefont {Pritchard},\ and\ \citenamefont {Shaffer}}]{Adams19jpbamop}%
  \BibitemOpen
  \bibfield  {author} {\bibinfo {author} {\bibfnamefont {C.~S.}\ \bibnamefont
  {Adams}}, \bibinfo {author} {\bibfnamefont {J.~D.}\ \bibnamefont
  {Pritchard}},\ and\ \bibinfo {author} {\bibfnamefont {J.~P.}\ \bibnamefont
  {Shaffer}},\ }\href {https://doi.org/10.1088/1361-6455/ab52ef} {\bibfield
  {journal} {\bibinfo  {journal} {J. Phys. B: At. Mol. Opt. Phys.}\ }\textbf
  {\bibinfo {volume} {53}},\ \bibinfo {pages} {012002} (\bibinfo {year}
  {2019})}\BibitemShut {NoStop}%
\bibitem [{\citenamefont {Araneda}\ \emph {et~al.}(2020)\citenamefont
  {Araneda}, \citenamefont {Cerchiari}, \citenamefont {Higginbottom},
  \citenamefont {Holz}, \citenamefont {Lakhmanskiy}, \citenamefont {Ob{\v
  s}il}, \citenamefont {Colombe},\ and\ \citenamefont {Blatt}}]{Araneda20rosi}%
  \BibitemOpen
  \bibfield  {author} {\bibinfo {author} {\bibfnamefont {G.}~\bibnamefont
  {Araneda}}, \bibinfo {author} {\bibfnamefont {G.}~\bibnamefont {Cerchiari}},
  \bibinfo {author} {\bibfnamefont {D.~B.}\ \bibnamefont {Higginbottom}},
  \bibinfo {author} {\bibfnamefont {P.~C.}\ \bibnamefont {Holz}}, \bibinfo
  {author} {\bibfnamefont {K.}~\bibnamefont {Lakhmanskiy}}, \bibinfo {author}
  {\bibfnamefont {P.}~\bibnamefont {Ob{\v s}il}}, \bibinfo {author}
  {\bibfnamefont {Y.}~\bibnamefont {Colombe}},\ and\ \bibinfo {author}
  {\bibfnamefont {R.}~\bibnamefont {Blatt}},\ }\href
  {https://doi.org/10.1063/5.0020661} {\bibfield  {journal} {\bibinfo
  {journal} {Review of Scientific Instruments}\ }\textbf {\bibinfo {volume}
  {91}},\ \bibinfo {pages} {113201} (\bibinfo {year} {2020})}\BibitemShut
  {NoStop}%
\bibitem [{\citenamefont {Bergmann}\ \emph {et~al.}(2019)\citenamefont
  {Bergmann}, \citenamefont {N{\"a}gerl}, \citenamefont {Panda}, \citenamefont
  {Gabrielse}, \citenamefont {Miloglyadov}, \citenamefont {Quack},
  \citenamefont {Seyfang}, \citenamefont {Wichmann}, \citenamefont {Ospelkaus},
  \citenamefont {Kuhn}, \citenamefont {Longhi}, \citenamefont {Szameit},
  \citenamefont {Pirro}, \citenamefont {Hillebrands}, \citenamefont {Zhu},
  \citenamefont {Zhu}, \citenamefont {Drewsen}, \citenamefont {Hensinger},
  \citenamefont {Weidt}, \citenamefont {Halfmann}, \citenamefont {Wang},
  \citenamefont {Paraoanu}, \citenamefont {Vitanov}, \citenamefont {Mompart},
  \citenamefont {Busch}, \citenamefont {Barnum}, \citenamefont {Grimes},
  \citenamefont {Field}, \citenamefont {Raizen}, \citenamefont {Narevicius},
  \citenamefont {Auzinsh}, \citenamefont {Budker}, \citenamefont {P{\'a}lffy},\
  and\ \citenamefont {Keitel}}]{bergmann2019stirap-review}%
  \BibitemOpen
  \bibfield  {author} {\bibinfo {author} {\bibfnamefont {K.}~\bibnamefont
  {Bergmann}}, \bibinfo {author} {\bibfnamefont {H.-C.}\ \bibnamefont
  {N{\"a}gerl}}, \bibinfo {author} {\bibfnamefont {C.}~\bibnamefont {Panda}},
  \bibinfo {author} {\bibfnamefont {G.}~\bibnamefont {Gabrielse}}, \bibinfo
  {author} {\bibfnamefont {E.}~\bibnamefont {Miloglyadov}}, \bibinfo {author}
  {\bibfnamefont {M.}~\bibnamefont {Quack}}, \bibinfo {author} {\bibfnamefont
  {G.}~\bibnamefont {Seyfang}}, \bibinfo {author} {\bibfnamefont
  {G.}~\bibnamefont {Wichmann}}, \bibinfo {author} {\bibfnamefont
  {S.}~\bibnamefont {Ospelkaus}}, \bibinfo {author} {\bibfnamefont
  {A.}~\bibnamefont {Kuhn}}, \bibinfo {author} {\bibfnamefont {S.}~\bibnamefont
  {Longhi}}, \bibinfo {author} {\bibfnamefont {A.}~\bibnamefont {Szameit}},
  \bibinfo {author} {\bibfnamefont {P.}~\bibnamefont {Pirro}}, \bibinfo
  {author} {\bibfnamefont {B.}~\bibnamefont {Hillebrands}}, \bibinfo {author}
  {\bibfnamefont {X.-F.}\ \bibnamefont {Zhu}}, \bibinfo {author} {\bibfnamefont
  {J.}~\bibnamefont {Zhu}}, \bibinfo {author} {\bibfnamefont {M.}~\bibnamefont
  {Drewsen}}, \bibinfo {author} {\bibfnamefont {W.~K.}\ \bibnamefont
  {Hensinger}}, \bibinfo {author} {\bibfnamefont {S.}~\bibnamefont {Weidt}},
  \bibinfo {author} {\bibfnamefont {T.}~\bibnamefont {Halfmann}}, \bibinfo
  {author} {\bibfnamefont {H.-L.}\ \bibnamefont {Wang}}, \bibinfo {author}
  {\bibfnamefont {G.~S.}\ \bibnamefont {Paraoanu}}, \bibinfo {author}
  {\bibfnamefont {N.~V.}\ \bibnamefont {Vitanov}}, \bibinfo {author}
  {\bibfnamefont {J.}~\bibnamefont {Mompart}}, \bibinfo {author} {\bibfnamefont
  {T.}~\bibnamefont {Busch}}, \bibinfo {author} {\bibfnamefont {T.~J.}\
  \bibnamefont {Barnum}}, \bibinfo {author} {\bibfnamefont {D.~D.}\
  \bibnamefont {Grimes}}, \bibinfo {author} {\bibfnamefont {R.~W.}\
  \bibnamefont {Field}}, \bibinfo {author} {\bibfnamefont {M.~G.}\ \bibnamefont
  {Raizen}}, \bibinfo {author} {\bibfnamefont {E.}~\bibnamefont {Narevicius}},
  \bibinfo {author} {\bibfnamefont {M.}~\bibnamefont {Auzinsh}}, \bibinfo
  {author} {\bibfnamefont {D.}~\bibnamefont {Budker}}, \bibinfo {author}
  {\bibfnamefont {A.}~\bibnamefont {P{\'a}lffy}},\ and\ \bibinfo {author}
  {\bibfnamefont {C.~H.}\ \bibnamefont {Keitel}},\ }\href
  {https://doi.org/10.1088/1361-6455/ab3995} {\bibfield  {journal} {\bibinfo
  {journal} {J. Phys. B: At. Mol. Opt. Phys.}\ }\textbf {\bibinfo {volume}
  {52}},\ \bibinfo {pages} {202001} (\bibinfo {year} {2019})}\BibitemShut
  {NoStop}%
\bibitem [{\citenamefont {Ghadimi}\ \emph {et~al.}(2017)\citenamefont
  {Ghadimi}, \citenamefont {Bl{\=u}ms}, \citenamefont {Norton}, \citenamefont
  {Fisher}, \citenamefont {Connell}, \citenamefont {Amini}, \citenamefont
  {Volin}, \citenamefont {Hayden}, \citenamefont {Pai}, \citenamefont
  {Kielpinski}, \citenamefont {Lobino},\ and\ \citenamefont
  {Streed}}]{Ghadimi17nqi}%
  \BibitemOpen
  \bibfield  {author} {\bibinfo {author} {\bibfnamefont {M.}~\bibnamefont
  {Ghadimi}}, \bibinfo {author} {\bibfnamefont {V.}~\bibnamefont {Bl{\=u}ms}},
  \bibinfo {author} {\bibfnamefont {B.~G.}\ \bibnamefont {Norton}}, \bibinfo
  {author} {\bibfnamefont {P.~M.}\ \bibnamefont {Fisher}}, \bibinfo {author}
  {\bibfnamefont {S.~C.}\ \bibnamefont {Connell}}, \bibinfo {author}
  {\bibfnamefont {J.~M.}\ \bibnamefont {Amini}}, \bibinfo {author}
  {\bibfnamefont {C.}~\bibnamefont {Volin}}, \bibinfo {author} {\bibfnamefont
  {H.}~\bibnamefont {Hayden}}, \bibinfo {author} {\bibfnamefont {C.-S.}\
  \bibnamefont {Pai}}, \bibinfo {author} {\bibfnamefont {D.}~\bibnamefont
  {Kielpinski}}, \bibinfo {author} {\bibfnamefont {M.}~\bibnamefont {Lobino}},\
  and\ \bibinfo {author} {\bibfnamefont {E.~W.}\ \bibnamefont {Streed}},\
  }\href {https://doi.org/10.1038/s41534-017-0006-6} {\bibfield  {journal}
  {\bibinfo  {journal} {npj Quantum Inf}\ }\textbf {\bibinfo {volume} {3}},\
  \bibinfo {pages} {1} (\bibinfo {year} {2017})}\BibitemShut {NoStop}%
\bibitem [{\citenamefont {Juliano~Martins}\ \emph {et~al.}(2022)\citenamefont
  {Juliano~Martins}, \citenamefont {Marinov}, \citenamefont {Youssef},
  \citenamefont {Kyrou}, \citenamefont {Joubert}, \citenamefont {Colmagro},
  \citenamefont {G{\^a}t{\'e}}, \citenamefont {Turbil}, \citenamefont {Coulon},
  \citenamefont {Turover}, \citenamefont {Khadir}, \citenamefont {Giudici},
  \citenamefont {Klitis}, \citenamefont {Sorel},\ and\ \citenamefont
  {Genevet}}]{JulianoMartins22nc}%
  \BibitemOpen
  \bibfield  {author} {\bibinfo {author} {\bibfnamefont {R.}~\bibnamefont
  {Juliano~Martins}}, \bibinfo {author} {\bibfnamefont {E.}~\bibnamefont
  {Marinov}}, \bibinfo {author} {\bibfnamefont {M.~A.~B.}\ \bibnamefont
  {Youssef}}, \bibinfo {author} {\bibfnamefont {C.}~\bibnamefont {Kyrou}},
  \bibinfo {author} {\bibfnamefont {M.}~\bibnamefont {Joubert}}, \bibinfo
  {author} {\bibfnamefont {C.}~\bibnamefont {Colmagro}}, \bibinfo {author}
  {\bibfnamefont {V.}~\bibnamefont {G{\^a}t{\'e}}}, \bibinfo {author}
  {\bibfnamefont {C.}~\bibnamefont {Turbil}}, \bibinfo {author} {\bibfnamefont
  {P.-M.}\ \bibnamefont {Coulon}}, \bibinfo {author} {\bibfnamefont
  {D.}~\bibnamefont {Turover}}, \bibinfo {author} {\bibfnamefont
  {S.}~\bibnamefont {Khadir}}, \bibinfo {author} {\bibfnamefont
  {M.}~\bibnamefont {Giudici}}, \bibinfo {author} {\bibfnamefont
  {C.}~\bibnamefont {Klitis}}, \bibinfo {author} {\bibfnamefont
  {M.}~\bibnamefont {Sorel}},\ and\ \bibinfo {author} {\bibfnamefont
  {P.}~\bibnamefont {Genevet}},\ }\href
  {https://doi.org/10.1038/s41467-022-33450-2} {\bibfield  {journal} {\bibinfo
  {journal} {Nat Commun}\ }\textbf {\bibinfo {volume} {13}},\ \bibinfo {pages}
  {5724} (\bibinfo {year} {2022})}\BibitemShut {NoStop}%
\bibitem [{\citenamefont {Kassa}(2017)}]{Kassa17Ths}%
  \BibitemOpen
  \bibfield  {author} {\bibinfo {author} {\bibfnamefont {E.}~\bibnamefont
  {Kassa}},\ }\emph {\bibinfo {title} {Single Ion Coupled to a High-Finesse
  Optical Fibre Cavity for {{cQED}} in the Strong Coupling Regime}},\
  \href@noop {} {\bibinfo {type} {Doctoral}},\ \bibinfo  {school} {University
  of Sussex} (\bibinfo {year} {2017})\BibitemShut {NoStop}%
\bibitem [{\citenamefont {Noek}\ \emph {et~al.}(2013)\citenamefont {Noek},
  \citenamefont {Vrijsen}, \citenamefont {Gaultney}, \citenamefont {Mount},
  \citenamefont {Kim}, \citenamefont {Maunz},\ and\ \citenamefont
  {Kim}}]{Noek13ol}%
  \BibitemOpen
  \bibfield  {author} {\bibinfo {author} {\bibfnamefont {R.}~\bibnamefont
  {Noek}}, \bibinfo {author} {\bibfnamefont {G.}~\bibnamefont {Vrijsen}},
  \bibinfo {author} {\bibfnamefont {D.}~\bibnamefont {Gaultney}}, \bibinfo
  {author} {\bibfnamefont {E.}~\bibnamefont {Mount}}, \bibinfo {author}
  {\bibfnamefont {T.}~\bibnamefont {Kim}}, \bibinfo {author} {\bibfnamefont
  {P.}~\bibnamefont {Maunz}},\ and\ \bibinfo {author} {\bibfnamefont
  {J.}~\bibnamefont {Kim}},\ }\href {https://doi.org/10.1364/OL.38.004735}
  {\bibfield  {journal} {\bibinfo  {journal} {Opt. Lett.}\ }\textbf {\bibinfo
  {volume} {38}},\ \bibinfo {pages} {4735} (\bibinfo {year}
  {2013})}\BibitemShut {NoStop}%
\bibitem [{\citenamefont {Northeast}\ \emph {et~al.}(2021)\citenamefont
  {Northeast}, \citenamefont {Dalacu}, \citenamefont {Weber}, \citenamefont
  {Phoenix}, \citenamefont {Lapointe}, \citenamefont {Aers}, \citenamefont
  {Poole},\ and\ \citenamefont {Williams}}]{Northeast21sr}%
  \BibitemOpen
  \bibfield  {author} {\bibinfo {author} {\bibfnamefont {D.~B.}\ \bibnamefont
  {Northeast}}, \bibinfo {author} {\bibfnamefont {D.}~\bibnamefont {Dalacu}},
  \bibinfo {author} {\bibfnamefont {J.~F.}\ \bibnamefont {Weber}}, \bibinfo
  {author} {\bibfnamefont {J.}~\bibnamefont {Phoenix}}, \bibinfo {author}
  {\bibfnamefont {J.}~\bibnamefont {Lapointe}}, \bibinfo {author}
  {\bibfnamefont {G.~C.}\ \bibnamefont {Aers}}, \bibinfo {author}
  {\bibfnamefont {P.~J.}\ \bibnamefont {Poole}},\ and\ \bibinfo {author}
  {\bibfnamefont {R.~L.}\ \bibnamefont {Williams}},\ }\href
  {https://doi.org/10.1038/s41598-021-02287-y} {\bibfield  {journal} {\bibinfo
  {journal} {Sci Rep}\ }\textbf {\bibinfo {volume} {11}},\ \bibinfo {pages}
  {22878} (\bibinfo {year} {2021})}\BibitemShut {NoStop}%
\bibitem [{\citenamefont {Park}\ \emph {et~al.}(2019)\citenamefont {Park},
  \citenamefont {Zhang}, \citenamefont {She}, \citenamefont {Chen},
  \citenamefont {Lin}, \citenamefont {Yousef}, \citenamefont {Cheng},\ and\
  \citenamefont {Capasso}}]{park-capasso2019nla}%
  \BibitemOpen
  \bibfield  {author} {\bibinfo {author} {\bibfnamefont {J.-S.}\ \bibnamefont
  {Park}}, \bibinfo {author} {\bibfnamefont {S.}~\bibnamefont {Zhang}},
  \bibinfo {author} {\bibfnamefont {A.}~\bibnamefont {She}}, \bibinfo {author}
  {\bibfnamefont {W.~T.}\ \bibnamefont {Chen}}, \bibinfo {author}
  {\bibfnamefont {P.}~\bibnamefont {Lin}}, \bibinfo {author} {\bibfnamefont
  {K.~M.~A.}\ \bibnamefont {Yousef}}, \bibinfo {author} {\bibfnamefont {J.-X.}\
  \bibnamefont {Cheng}},\ and\ \bibinfo {author} {\bibfnamefont
  {F.}~\bibnamefont {Capasso}},\ }\href
  {https://doi.org/10.1021/acs.nanolett.9b03333} {\bibfield  {journal}
  {\bibinfo  {journal} {Nano Lett.}\ }\textbf {\bibinfo {volume} {19}},\
  \bibinfo {pages} {8673} (\bibinfo {year} {2019})}\BibitemShut {NoStop}%
\bibitem [{\citenamefont {Ruelle}\ \emph {et~al.}(2022)\citenamefont {Ruelle},
  \citenamefont {Jaeger}, \citenamefont {Fogliano}, \citenamefont {Braakman},\
  and\ \citenamefont {Poggio}}]{Ruelle22rosi}%
  \BibitemOpen
  \bibfield  {author} {\bibinfo {author} {\bibfnamefont {T.}~\bibnamefont
  {Ruelle}}, \bibinfo {author} {\bibfnamefont {D.}~\bibnamefont {Jaeger}},
  \bibinfo {author} {\bibfnamefont {F.}~\bibnamefont {Fogliano}}, \bibinfo
  {author} {\bibfnamefont {F.}~\bibnamefont {Braakman}},\ and\ \bibinfo
  {author} {\bibfnamefont {M.}~\bibnamefont {Poggio}},\ }\href
  {https://doi.org/10.1063/5.0098140} {\bibfield  {journal} {\bibinfo
  {journal} {Review of Scientific Instruments}\ }\textbf {\bibinfo {volume}
  {93}},\ \bibinfo {pages} {095003} (\bibinfo {year} {2022})}\BibitemShut
  {NoStop}%
\bibitem [{\citenamefont {Shen}\ \emph {et~al.}(2020)\citenamefont {Shen},
  \citenamefont {Chen}, \citenamefont {Wu}, \citenamefont {Dong}, \citenamefont
  {Cui}, \citenamefont {You},\ and\ \citenamefont {Tey}}]{Shen20rosi}%
  \BibitemOpen
  \bibfield  {author} {\bibinfo {author} {\bibfnamefont {C.}~\bibnamefont
  {Shen}}, \bibinfo {author} {\bibfnamefont {C.}~\bibnamefont {Chen}}, \bibinfo
  {author} {\bibfnamefont {X.-L.}\ \bibnamefont {Wu}}, \bibinfo {author}
  {\bibfnamefont {S.}~\bibnamefont {Dong}}, \bibinfo {author} {\bibfnamefont
  {Y.}~\bibnamefont {Cui}}, \bibinfo {author} {\bibfnamefont {L.}~\bibnamefont
  {You}},\ and\ \bibinfo {author} {\bibfnamefont {M.~K.}\ \bibnamefont {Tey}},\
  }\href {https://doi.org/10.1063/5.0006026} {\bibfield  {journal} {\bibinfo
  {journal} {Review of Scientific Instruments}\ }\textbf {\bibinfo {volume}
  {91}},\ \bibinfo {pages} {063202} (\bibinfo {year} {2020})}\BibitemShut
  {NoStop}%
\bibitem [{\citenamefont {Stephenson}(2019)}]{Stephenson19Ths}%
  \BibitemOpen
  \bibfield  {author} {\bibinfo {author} {\bibfnamefont {L.}~\bibnamefont
  {Stephenson}},\ }\emph {\bibinfo {title} {Entanglement between Nodes of a
  Quantum Network}},\ \href@noop {} {\bibinfo {type}
  {{{http://purl.org/dc/dcmitype/Text}}}},\ \bibinfo  {school} {University of
  Oxford} (\bibinfo {year} {2019})\BibitemShut {NoStop}%
\bibitem [{\citenamefont {Stephenson}\ \emph {et~al.}(2020)\citenamefont
  {Stephenson}, \citenamefont {Nadlinger}, \citenamefont {Nichol},
  \citenamefont {An}, \citenamefont {Drmota}, \citenamefont {Ballance},
  \citenamefont {Thirumalai}, \citenamefont {Goodwin}, \citenamefont {Lucas},\
  and\ \citenamefont {Ballance}}]{Stephenson20prl}%
  \BibitemOpen
  \bibfield  {author} {\bibinfo {author} {\bibfnamefont {L.~J.}\ \bibnamefont
  {Stephenson}}, \bibinfo {author} {\bibfnamefont {D.~P.}\ \bibnamefont
  {Nadlinger}}, \bibinfo {author} {\bibfnamefont {B.~C.}\ \bibnamefont
  {Nichol}}, \bibinfo {author} {\bibfnamefont {S.}~\bibnamefont {An}}, \bibinfo
  {author} {\bibfnamefont {P.}~\bibnamefont {Drmota}}, \bibinfo {author}
  {\bibfnamefont {T.~G.}\ \bibnamefont {Ballance}}, \bibinfo {author}
  {\bibfnamefont {K.}~\bibnamefont {Thirumalai}}, \bibinfo {author}
  {\bibfnamefont {J.~F.}\ \bibnamefont {Goodwin}}, \bibinfo {author}
  {\bibfnamefont {D.~M.}\ \bibnamefont {Lucas}},\ and\ \bibinfo {author}
  {\bibfnamefont {C.~J.}\ \bibnamefont {Ballance}},\ }\href
  {https://doi.org/10.1103/PhysRevLett.124.110501} {\bibfield  {journal}
  {\bibinfo  {journal} {Phys. Rev. Lett.}\ }\textbf {\bibinfo {volume} {124}},\
  \bibinfo {pages} {110501} (\bibinfo {year} {2020})}\BibitemShut {NoStop}%
\bibitem [{\citenamefont {Wang}\ \emph {et~al.}(2020)\citenamefont {Wang},
  \citenamefont {Goham}, \citenamefont {Laugharn},\ and\ \citenamefont
  {Britton}}]{Wang20oq2c2pq}%
  \BibitemOpen
  \bibfield  {author} {\bibinfo {author} {\bibfnamefont {W.}~\bibnamefont
  {Wang}}, \bibinfo {author} {\bibfnamefont {C.}~\bibnamefont {Goham}},
  \bibinfo {author} {\bibfnamefont {A.}~\bibnamefont {Laugharn}},\ and\
  \bibinfo {author} {\bibfnamefont {J.~W.}\ \bibnamefont {Britton}},\ }in\
  \href {https://doi.org/10.1364/QUANTUM.2020.QW6A.12} {\emph {\bibinfo
  {booktitle} {{{OSA Quantum}} 2.0 {{Conference}} (2020), Paper {{QW6A}}.12}}}\
  (\bibinfo  {publisher} {Optica Publishing Group},\ \bibinfo {year} {2020})\
  p.\ \bibinfo {pages} {QW6A.12}\BibitemShut {NoStop}%
\bibitem [{\citenamefont {Xiao}\ \emph {et~al.}(2021)\citenamefont {Xiao},
  \citenamefont {Wang}, \citenamefont {Wang}, \citenamefont {Lee},
  \citenamefont {Kananen},\ and\ \citenamefont {Gu}}]{Xiao21j}%
  \BibitemOpen
  \bibfield  {author} {\bibinfo {author} {\bibfnamefont {Y.}~\bibnamefont
  {Xiao}}, \bibinfo {author} {\bibfnamefont {Z.}~\bibnamefont {Wang}}, \bibinfo
  {author} {\bibfnamefont {F.}~\bibnamefont {Wang}}, \bibinfo {author}
  {\bibfnamefont {H.}~\bibnamefont {Lee}}, \bibinfo {author} {\bibfnamefont
  {T.}~\bibnamefont {Kananen}},\ and\ \bibinfo {author} {\bibfnamefont
  {T.}~\bibnamefont {Gu}},\ }\href {https://doi.org/10.1117/1.JOM.1.2.024001}
  {\bibfield  {journal} {\bibinfo  {journal} {JOM}\ }\textbf {\bibinfo {volume}
  {1}},\ \bibinfo {pages} {024001} (\bibinfo {year} {2021})}\BibitemShut
  {NoStop}%
\bibitem [{\citenamefont {Young}\ \emph {et~al.}(2022)\citenamefont {Young},
  \citenamefont {Safari}, \citenamefont {Huft}, \citenamefont {Zhang},
  \citenamefont {Oh}, \citenamefont {Chinnarasu},\ and\ \citenamefont
  {Saffman}}]{Young22apb}%
  \BibitemOpen
  \bibfield  {author} {\bibinfo {author} {\bibfnamefont {C.~B.}\ \bibnamefont
  {Young}}, \bibinfo {author} {\bibfnamefont {A.}~\bibnamefont {Safari}},
  \bibinfo {author} {\bibfnamefont {P.}~\bibnamefont {Huft}}, \bibinfo {author}
  {\bibfnamefont {J.}~\bibnamefont {Zhang}}, \bibinfo {author} {\bibfnamefont
  {E.}~\bibnamefont {Oh}}, \bibinfo {author} {\bibfnamefont {R.}~\bibnamefont
  {Chinnarasu}},\ and\ \bibinfo {author} {\bibfnamefont {M.}~\bibnamefont
  {Saffman}},\ }\href {https://doi.org/10.1007/s00340-022-07865-0} {\bibfield
  {journal} {\bibinfo  {journal} {Appl. Phys. B}\ }\textbf {\bibinfo {volume}
  {128}},\ \bibinfo {pages} {151} (\bibinfo {year} {2022})}\BibitemShut
  {NoStop}%
\end{thebibliography}%

\end{document}